\documentclass[onecolumn]{aa} 
\usepackage{booktabs}
\usepackage{graphicx}
\usepackage{amsmath}
\usepackage[normalem]{ulem}
\usepackage[nolist]{acronym}
\usepackage[separate-uncertainty=true]{siunitx}
\usepackage{lineno}

\usepackage{txfonts}
\usepackage{hyperref}

\usepackage[nolist]{acronym}
\acrodef{StEP}[SEP]{Stellar Energetic Particle}
\acrodef{SCR}[SCR]{Stellar Cosmic Ray}
\acrodef{GCR}[GCR]{Galactic Cosmic Ray}
\acrodef{JWST}[JWST]{James Webb Space Telescope}
\acrodef{E-ELT}[E-ELT]{European Extremely Large Telescope}
\acrodef{HabEX}[HabEX]{Habitable Exoplanet Observatory}
\acrodef{LUVOIR}[LUVOIR]{Large UV/Optical/IR Surveyor}
\acrodef{ISM}[ISM]{Interstellar Medium}
\acrodef{LIS}[LIS]{Local Interstellar Spectrum}
\acrodef{FFA}[FFA]{Force-Field Approximation}
\acrodef{GLE}[GLE]{Ground Level Enhancement}
\acrodef{PSD}[PSD]{peak size distribution function}
\acrodef{TOA}[TOA]{top of the atmosphere}
\acrodef{AtRIS}[AtRIS]{\textbf{At}mospheric \textbf{R}adiation \textbf{I}nteraction \textbf{S}imulator}
\acrodef{KIT}[KIT]{Karlsruher Institute of Technology}

\usepackage{xcolor}
\definecolor{ForestGreen}{HTML}{228B22}
\definecolor{Mulberry}{HTML}{C54B8C}
\definecolor{Dandelion}{HTML}{FFCC99}
\definecolor{RoyalBlue}{HTML}{4169E1}
\definecolor{Skobeloff}{HTML}{007474}

\begin{document} 



\title{A New Model Suite to Determine the Influence of Cosmic Rays on (Exo)planetary Atmospheric Biosignatures}
\subtitle{Validation based on Modern Earth}

   \author{Konstantin Herbst\inst{1}
          \and
          John Lee Grenfell\inst{2}
          \and
          Miriam Sinnhuber\inst{3}
          \and
          Heike Rauer\inst{2}
          \and
          Bernd Heber\inst{1}
          \and
          Sa\v{s}a Banjac\inst{1}
          \and
          Markus Scheucher\inst{4}
          \and
          Vanessa Schmidt\inst{3}
          \and
          Stefanie Gebauer\inst{2}
          \and
          Ralph Lehmann\inst{5}
          \and 
          Franz Schreier\inst{6}
          }

   \institute{Institut f\"ur Experimentelle und Angewandte Physik (IEAP), Christian-Albrechts-Universit\"at zu Kiel (CAU), 24118 Kiel, Germany\\
              \email{herbst@physik.uni-kiel.de}
         \and
             Institut f\"ur Planetenforschung (PF), Deutsches Zentrum f\"ur  Luft- und Raumfahrt (DLR), Rutherfordstr. 2, 12489 Berlin, Germany
         \and
             Institut f\"ur Meteorologie und Klimaforschung, Karlsruher Institut für Technologie (KIT), Hermann-von Helmholtz Platz 1, 76344 Eggenstein-Leopoldshafen, Germany
         \and
             Zentrum f\"ur Astronomie und Astrophysik, Technische Universit\"at Berlin (TUB), Hardenbergstrasse 36, 10623 Berlin, Germany
        \and
            Alfred Wegener Institut, Helmholtz Zentrum f\"ur Polar- und Meeresforschung, Telegrafenberg A45, 14473 Potsdam, Germany
        \and
             Institut f\"ur Methodik der Fernerkundung, Deutsches Zentrum f\"ur Luft- und Raumfahrt (DLR), 82234 Oberpfaffenhofen, Germany}
\date{}
\authorrunning{Herbst et al.}
\titlerunning{Influence of Cosmic Rays on Exoplanetary Atmospheric Biosignatures}
  \abstract
 %
{The first opportunity to detect indications for life outside the Solar System may be provided already within the next decade with upcoming missions such as the James Webb Space Telescope (JWST), the European Extremely Large Telescope (E-ELT) and/or the Atmospheric Remote-sensing Infrared Exoplanet Large-survey (ARIEL) mission, searching for atmospheric biosignatures on planets in the habitable zone of cool K- and M-stars. Nevertheless, their harsh stellar radiation and particle environment could lead to photochemical loss of atmospheric biosignatures.}
{We aim to study the influence of cosmic rays on exoplanetary atmospheric biosignatures and the radiation environment considering feedbacks between energetic particle precipitation, climate, atmospheric ionization, neutral and ion chemistry, and secondary particle generation.}
{We describe newly-combined state-of-the-art modeling tools to study the impact of the radiation and particle environment, in particular of cosmic rays, on atmospheric particle interaction, the influence on the atmospheric chemistry and the climate-chemistry coupling in a self-consistent model suite. To this end, models like the Atmospheric Radiation Interaction Simulator (AtRIS), the Exoplanetary Terrestrial Ion Chemistry model (ExoTIC), and the updated coupled climate-chemistry model are combined.}
{Besides comparing our results to Earth-bound measurements, we investigate the ozone-production and -loss cycles as well as the atmospheric radiation dose profiles during quiescent solar periods and during the strong solar energetic particle event of February 23, 1956. Further, the scenario-dependent terrestrial transit spectra, as seen by the NIR-Spec infrared spectrometer onboard the JWST, are modeled. Amongst others, we find that the comparatively weak solar event (i) drastically increases the spectral signal of HNO$_3$, while (ii) significantly suppressing the spectral feature of ozone. Because of the slow recovery after such events, the latter indicates that ozone might not be a good biomarker for planets orbiting stars with high flaring rates.
}
   {}
\keywords{cosmic rays -- planets and satellites: terrestrial planets -- planets and satellites: atmospheres -- biosignatures -- astrobiology}

\maketitle
\section{Introduction}

Cool stars such as K- and M-dwarfs are prime targets in the search for habitable Earth-like planets due to high occurrence rates in the solar neighborhood, small radii (leading to favorable planet-to-star contrast ratios) and rather close-in habitable zones (hence short intervals between transit events)   \citep[see, e.g.][and references therein]{Gillon-etal-2017,Dittmann-etal-2017}. Although recent results suggest that rocky planets in the habitable zone (HZ) of these host-stars could retain liquid water on their surface \citep[][]{Noack-etal-2017}, their presence within the HZ does not automatically imply that they are habitable \citep[see, e.g.][]{Lammer-etal-2009}. The close-in HZ and a possibly weakened planetary magnetic field due to potential tidal-locking \citep[see, e.g.,][]{Scalo-etal-2007} suggests that these worlds could be subject to a large input from stellar UV photons and energetic particles modulated by their stellar wind. There are essentially two types of energetic particles: \acp{GCR} from, e.g., supernova remnants \citep[e.g.][and references therein]{Buesching-etal-2005} and \acp{SCR}, which arise from stellar activity \citep[see, e.g.,][and references therein]{Buccino-etal-2007,France-etal-2013}. The astrosphere, i.e., the region around a star is filled by plasma of stellar origin \citep[see for example][]{Wood-etal-2005}, protects the (exo)planet against the isotropically distributed \acp{GCR}. The extension of the astrosphere, and thus the amount of protection, is determined by the stellar wind. The (exo)planetary magnetosphere, the magnetic environment of the planet, protects against \acp{StEP} and the stellar wind. Factors such as the presence of stellar coronal mass ejections \citep{Khodachenko-etal-2007} can lead to the erosion of unprotected planetary atmospheres. Furthermore, extreme \ac{StEP} events may cause drastic radiation dose increases and, thus, may drastically impact life \citep[][]{Vidotto-etal-2013, doNascimento-etal-2016ApJ}.

Moreover, the different stellar instellation could lead to planetary surface conditions, which are very different from those on Earth in terms of cosmic ray (CR) flux, radiation, and climate. For these reasons, studying the influence of the stellar radiation and particle environment upon planetary conditions in a physically-consistent manner requires a suite of state-of-the-art modeling tools together with expertise from numerous interdisciplinary fields such as space and atmospheric physics and biogeochemical-climate modeling. Due to the complexity of the problem, limiting initial model studies to Earth-like exoplanets is quite common in literature. In reality, however, the potential range of, e.g., atmospheres, biospheres, and interiors is ill-constrained. Furthermore, due to their relatively close-in orbits, Earth-like exoplanets around K- and M-stars could be tidally-locked to their parent star. Consequently, the presence of sufficient atmospheric \citep[see e.g.][]{Joshi-etal-1997} or oceanic \citep[see, e.g.,][]{Hu-Yang-2014} mass for meridional transport of energy from the day to the night side may be seen as a precondition for habitability. Numerous numerical 3D studies for rocky exoplanets in the HZ of K- and M-stars have recently been published  \citep[see, e.g.,][]{Leconte-etal-2013, Shields-etal-2013,Yang-etal-2013, Godolt-etal-2015, Kopparapu-etal-2016}. Such studies are vital since potentially rocky exoplanets orbiting in the HZ of cool stars are already starting to be detected by combining the transit method (determining radius) with the radial velocity method (determining mass), and more are expected to be found by the upcoming space missions. Therefore, theoretical studies are required now to understand the expected range of atmospheres, interiors, climates and spectral signatures in the context of next-generation missions such as the \acl{JWST} \citep[\acs{JWST}, ][]{Kaltenegger-Traub-2009,Belu-etal-2011,Rauer-etal-2011} the \acl{E-ELT} \citep[\acs{E-ELT}, e.g. ][]{Snellen-etal-2013} as well as further-afield missions such as the \ac{HabEX} and the \ac{LUVOIR}. A summary of future exoplanet missions in an exoplanetary context can be found, for example in \citet{Fujii-etal-2018}. Recent publications by \citet{Schwietermann-etal-2018} and \citet{Grenfell-etal-2018} provide a comprehensive review of current biosignature science and review photochemical gas-phase biosignatures responses, respectively. \citet{Shields-etal-2016} recently reviewed the potential habitability of planets in the HZ of M-stars.

In summary, in order to study planetary habitability, detailed knowledge of (i) the astrospheric particle- and radiation environment, (ii) the atmospheric chemistry, as well as (iii) the climate-chemistry coupling is mandatory. Although cool stars are favored targets for investigating rocky exoplanets, the effect of strong stellar energetic particles upon potential atmospheric biosignatures is not well known.

Here we present our approach on building up a self-consistent simulation chain coupling state-of-the-art magnetospheric and atmospheric propagation and interaction models with atmospheric chemistry and climate models  in our study involving the Christian-Albrechts-Universtit\"at (CAU), the Deutsches Zentrum f\"ur Luft- und Raumfahrt (DLR) and the \ac{KIT} which is supported by the German Research Foundation (DFG).

The present study focuses on the validation of our model chain based on modern Earth as an "exoplanet" in the HZ of a G2V star during quiescent as well as constant flaring conditions. Future studies will focus on exoplanetary scenarios such as the potentially Earth-like exoplanet Proxima Centauri b.
\section{Scientific Background}
\subsection{Particle Propagation and Transport in Magnetospheres and Atmospheres}
The structure of an astrosphere strongly depends on the properties of both the hot and fully ionized stellar wind as well as the surrounding \ac{ISM}. Differences in the properties of the stellar wind and the local \ac{ISM} may thereby lead to a wide range of shapes for observable astrospheres \citep[see, e.g.,][]{Kobulnicky-etal-2016}. Simulations of astrospheres around hot stars most often make use of 1D or 2D (magneto-)hydrodynamical approaches \citep[see, e.g., ][]{Marle-etal-2014, Cos-etal-2016}. However, in order to estimate the radiation and cosmic ray particle field of G-, K- and M-stars, 3D-astrospheric modeling becomes inevitable. First studies based on the full-3D MHD code Cronos \citep[see, e.g,][]{Kissmann-etal-2008, Wiengarten-etal-2013} as well as the 3D kinetic MHD model by \citet{Izmodenov-Alexashov-2015} have been successfully used by, e.g., \citet{Scherer-etal-2015} and \citet{Katushkina-etal-2018} to model the astrospheres of hot O-type stars, respectively.

Moreover, the transport of \acp{GCR} through an astrosphere depends on its volume, structure and turbulent state, which are governed by, e.g., the stellar type, rotation rate, stellar wind dynamics, stellar activity as well as the magnetic field as inner boundary conditions and the \ac{LIS} as an outer boundary condition. In the case of the Sun, the latter, however, was only recently derived from measurements by the Voyager 1 spacecraft \citep[see, e.g.][]{Stone-etal-2013, Webber-McDonald-2013}, and several \ac{LIS} models exist in literature \citep[an overview of the models is given by][]{Herbst-etal-2017}. By treating the astrosphere as a heliospheric-analog, transport and modulation of \acp{GCR} can be investigated by applying Parker's transport equation \citep[see][]{Parker-1968}. 

\begin{figure}[t!]
\begin{center}
\includegraphics[width=0.7\columnwidth]{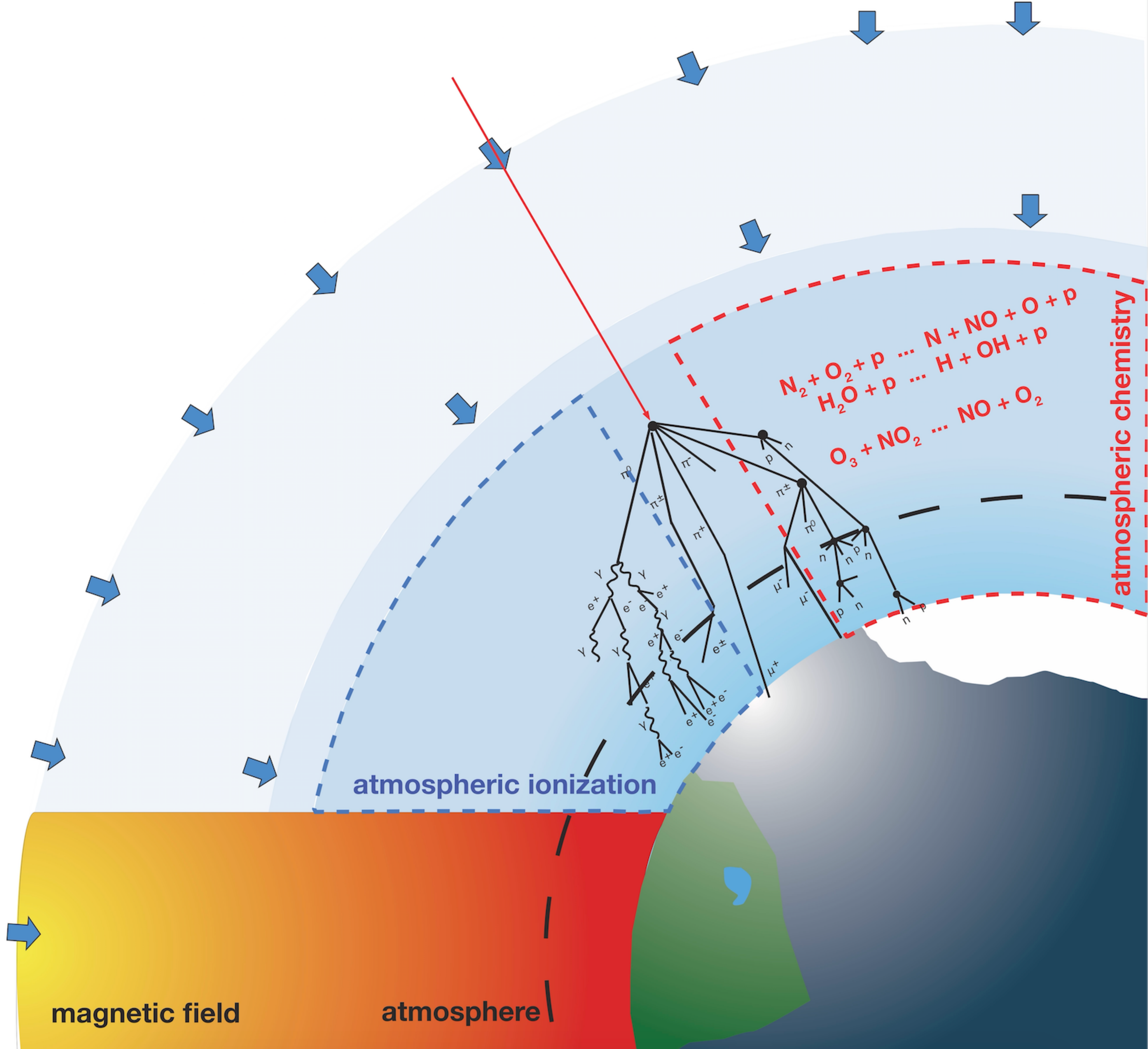}
  \caption{Processes induced by cosmic rays affecting atmospheric climate and ion-relevant chemistry of an Earth-like exoplanet.}
 \label{fig:1}
 \end{center}
 \end{figure}

An often-used approximation is the so-called \ac{FFA} which can be applied provided the planet is located well inside the astro- or heliosphere \citep{Caballero-Lopez-etal-2004, Moraal-2013}. This first-order approximation of the GCR flux at a distance of 1~AU is given by:
\begin{equation}\label{Eq.2}
J(E, \phi) = J_{\mathrm{LIS}}(E+\phi) \cdot \frac{E \cdot (E + 2 E_\mathrm{r})}{(E+\Phi) \cdot (E+\Phi+2E_\mathrm{r})},
\end{equation}
where $J_{\mathrm{LIS}}$ represents the unmodulated differential \ac{GCR} flux outside the heliosphere and $E_\mathrm{r}$ denotes the rest energy of the primary particle \citep{Moraal-2013}. The stellar modulation function is given by $\Phi$ = (Ze/A) $\phi$ (in units of GeV), with Z as the charge and A as the mass number of the particle, while $\phi$ represents the so-called modulation parameter, a measure of the stellar activity (in units of GV). High stellar activity, e.g., corresponds to strong modulation, i.e., high $\phi$ values and, thus, low GCR fluxes and vice versa. According to \citet{Caballero-Lopez-etal-2004}, $\phi$ can be derived via:
\begin{equation}
\phi(r) \equiv \int_{P}^{P_b(r, P)} \frac{\beta(P')\kappa_2(P')}{P'} dP' = \int_r^{r_b} \frac{v(r')}{3 \kappa_1(r')} dr',
\end{equation}
where $r_b$ represents the outer boundary of the astrosphere, $v$ the stellar wind velocity and $\kappa$ the diffusion coefficient. Thus, in theory, the \ac{FFA} can be applied to determine the GCR flux within any astrosphere. However, the approach, among others, neglects adiabatic cooling processes and can depend on additional parameters such as the magnetic field structure, particle drifts, stellar cycle, etc. \citep[][and references therein]{Gieseler-etal-2017}. Thus, an application of the \ac{FFA} to other stars requires knowledge of the astrospheric turbulence (represented by $\kappa$) as well as the stellar wind velocity. Although stellar $\kappa$ values are unknown (and thus usually replaced by corresponding heliospheric values), the stellar wind velocity can be calculated with stellar wind models \citep[see, e.g.,][]{Preusse-etal-2005, Vidotto-etal-2011}. The required input parameters, such as the stellar magnetic field strength, the effective stellar temperature as well as the mass flux, are frequently challenging to obtain, but can also be derived indirectly \citep[see, e.g.,][]{Preusse-etal-2006, Kopp-etal-2011}. Such models predict weaker winds, and thus lower $\phi$-values, for quiet K-stars (with its HZ centered at approximately 0.6~AU) but considerably stronger winds for active M-stars (assuming an HZ centered at approximately 0.2~AU) compared to our G2V star, the Sun. Consequently, the energetic particle flux at the orbit of a planet in the HZ around quiet K- stars may be dominated by the \ac{GCR} component, whereas the \ac{StEP} component is considerably more important for planets around active K- and M-stars due to high stellar activity and strong stellar winds.

Strong solar activity can also lead to particles that are being accelerated up to several GeV spiraling along the heliospheric magnetic field lines. Planets well-connected to these field lines will encounter an enhanced solar particle flux, which may be detected at the planetary surface of, for example, Earth and Mars \citep[see, e.g.,][]{Shea-Smart-2000}. These events are so-called \acp{GLE} and are also expected to occur in exoplanetary atmospheres. The incoming stellar cosmic ray flux on exoplanets in the HZ of cooler stars, however, currently is not observable. Therefore, \citet{Segura-etal-2010} proposed an indirect means to estimate the \ac{StEP} flux based on so-called  \acf{PSD} \citep[see, e.g.,][]{Belov-etal-2005} reflecting the power-law relation between peak X-ray flare intensities and peak proton fluxes. However, most recently, \citet{Herbst-etal-2019a} showed that estimating the proton fluxes around K- and M-stars is not as straightforward as previously thought. Based on a new Double-Power-Law function they, for the first time provide best and worst-case estimates of the stellar proton fluxes around K- and M-dwarf, indicating that previous studies may have underestimated the stellar particle fluxes. 

Besides, planetary magnetospheres act as a filter for low-energy particles. Thus, the CR flux at the \ac{TOA} depends on the magnitude and geometry of the planetary magnetic field \citep{Fichtner-etal-2012, Herbst-etal-2013}. However, because rocky exoplanets in the HZ around M-stars may be tidally-locked, weaker exoplanetary magnetic fields could be present \citep[see, e.g.,][]{Christensen-Aubert-2006, Griessmeier-etal-2009}. In addition, \citet{Cohen-etal-2014} suggested that the magnetosphere structure of habitable planets orbiting M-dwarfs could be widely different from that of the Earth, for example, with potentially strong Joule heating due to strong stellar winds impinging on the upper planetary atmosphere.

Nevertheless, an energetic charged particle entering the \ac{TOA} will lose its energy within an atmosphere mostly due to collisions, leading to an atmospheric ionization, dissociative ionization, excitation and dissociation of the most abundant species in the upper layers. Fig.~\ref{fig:1} shows schematically the incoming trajectory of a CR particle (solid red line) penetrating the Earth’s magnetic field. Besides ionizing the upper layers, interactions with the atmospheric constituents like nitrogen, oxygen or argon trigger the evolution of a secondary particle shower as well as associated photochemical effects (see below). The probability of an energetic CR particle colliding with atmospheric species increases with atmospheric depth. The resulting energetic secondary particles may suffer further interactions forming an atmospheric particle cascade \citep{Dorman-etal-2004}. A significant quantity describing the effect of energetic particles on life is the radiation dose \citep{Horneck-etal-2001, Zeitlin-etal-2011},
which depends on the particle species. Furthermore, electrons created in the secondary particle showers may lead to a breakup of molecular nitrogen and oxygen followed by ion and gas-phase chemistry that produces nitrogen and hydrogen oxides. These, in turn, can catalytically remove atmospheric ozone, hence lead to enhanced harmful UV radiation reaching the surface. Recent results for the impact of particle precipitation on the formation of secondary particles, ionization, and radiation dose in the terrestrial atmosphere have been discussed in e.g., \citet{Mironova-etal-2015} and \citet{banjac2018}.

A unified simulation code that can take into account the stellar radiation and particle environment, and model the particle interactions from the magnetosphere, through the ionosphere, down to the surface has been rather lacking in the literature. Recently, however, \citet{banjac2018} developed the Geant4-based \ac{AtRIS} code with which diverse planetary configurations can be realized. First validations on the planetary atmospheres of Venus \citep[thick CO$_2$-dominated atmosphere][]{Herbst-etal-2019b}, Earth \citep[ N$_2$-O$_2$-dominated atmosphere][]{banjac2018} and Mars \citep[thin CO$_2$-dominated atmosphere][]{Guo-etal-2019} have been performed.
\subsection{Cosmic-Ray Induced Ion Chemistry}
\begin{figure}[!t]
\begin{center}
\includegraphics[width=0.7\columnwidth]{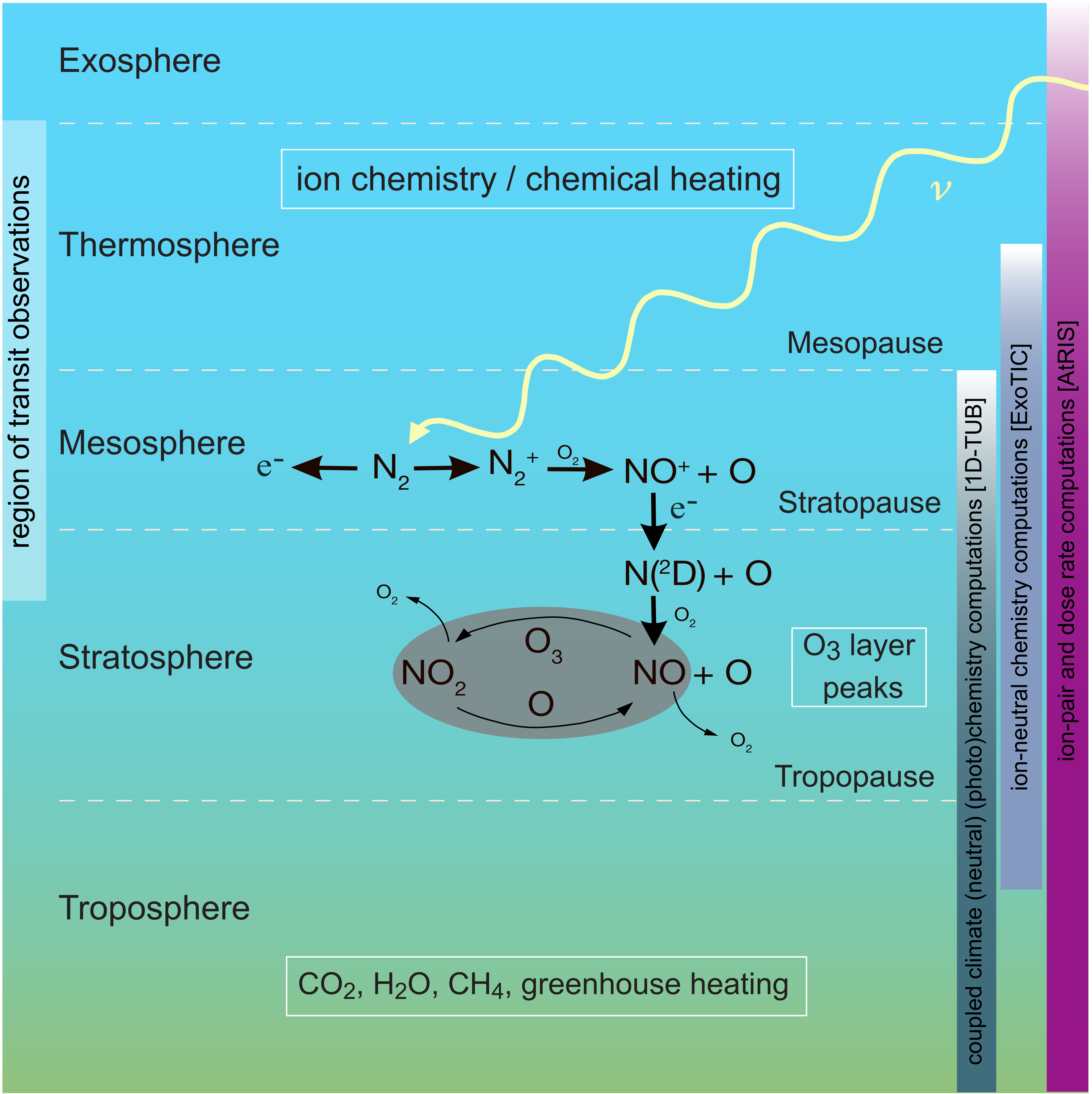}
  \caption{Schematic showing altitude-dependent chemical interactions within an Earth-like atmosphere. The highlighted altitudes show the composition- and wavelength-dependent region typically sampled by transmission spectroscopy. Besides, we highlighted the regions at which the models of our model suite (discussed in Section~\ref{sec:AtRIS_1DTUB_ExoTIC}) operate.}
 \label{fig:2}
\end{center}
 \end{figure}

Figure \ref{fig:2} shows schematically ion-related chemical processes induced by GCRs and SCRs. Energetic particles precipitating into the atmosphere collide with atmospheric constituents, leading to excitation, dissociation, and ionization of the most abundant species. In an N$_2$-O$_2$ atmosphere such as on present-day Earth, this leads to the production of, e.g., excited atomic states such as N($^2$D) and O($^1$D), or ions such as N$_2^+$, O$_2^+$, or O$^+$. In CO$_2$ dominated atmospheres, ions and excited-state dissociation products of CO$_2$ would likely be the primary products. The ionization of the atmosphere leads to chains of very fast ion-chemistry reactions, which in Earth-like atmospheres lead to the formation of neutral reactive radicals such as NO, H, and OH. These contribute to catalytic reaction chains which overall destroy ozone such as that shown, e.g., in Eqs.~(3) and (4):
\begin{eqnarray}
\mathrm{NO} + \mathrm{O}_3 &\longrightarrow& \mathrm{NO}_2 + \mathrm{O}_2\\
\mathrm{NO}_2 + \mathrm{O}(^3\mathrm{P}) &\longrightarrow& \mathrm{NO} + \mathrm{O}_2
\end{eqnarray}
Large ion clusters can form by incorporating molecules such as HNO$_3$, N$_2$O$_2$, or HCl, hence changing the partitioning of nitrogen and chlorine-containing species; this will also affect the ozone content. Since ozone is the main contributor to radiative heating of the stratosphere, this directly affects atmospheric temperatures and circulation. A summary of state of the art for Earth's atmosphere is provided, for example, in the review paper by \citet{Sinnhuber-etal-2012}. 

Ion-chemistry however, can also produce hydrocarbon haze from methane. This, in particular, is of importance for methane-rich atmospheres like those of Jupiter and Titan \citep[see, e.g.,][]{Vuitton-etal-2007, Lavvas-etal-2013, Hoerst-2017, Garcia-Munoz-2018}, respectively. In addition, methane-rich and $N_2$-dominated atmospheres can be seen as natural laboratories to study chemical evolution \citep[see, e.g.,][]{Kobayashi-etal-2017} and its importance for prebiotic chemistry \citep[see, e.g.,][]{Rimmer-Rugheimer-2019}.
%
\subsection{Atmospheric Biosignatures}
We provide here a short discussion on some of the atmospheric biosignatures commonly discussed in the exoplanetary literature. Note, that this list is not exhaustive.
\\
\\
\textbf{Ozone (O$_3$) -} around 90$\%$ of Earth’s O$_3$ is found in the stratosphere, the remaining ~10$\%$ lies in the troposphere. O$_3$ is produced mainly through the Chapman mechanism \citep{Chapman-1930} via O$_2$ photolysis in the Earth’s stratosphere and in the troposphere via the smog mechanism \citep{Haagen-Smit-1952}, which requires volatile organic compounds, nitrogen oxides and UV radiation. Thereby, O$_3$ is an indirect biosignature of life (i.e. the direct biosignature is O$_2$). Nevertheless, O$_3$ is often focused upon in exoplanetary biosignature studies due to strong spectral absorption and because its atmospheric abundance remains relatively constant over a wide range of O$_2$ abundances. O$_3$ sinks can be rather complex and include for example catalytic cycles involving, e.g., NO$_x$ and HO$_x$ \citep{Bates-Nicolet-1950, Crutzen-1970}. Photochemical modeling responses are discussed in Section \ref{sec:2}.
\\
\\
\textbf{Nitrous oxide (N$_2$O) -} in Earth’s atmosphere is formed almost exclusively as a by-product from the respiration of nitrifying and denitrifying bacteria as part of Earth’s nitrogen cycle \citep{Syakila-Kroeze-2011}. It has only very weak abiotic sources and, thus, is a good biosignature. However, its spectral features in modern Earth's atmosphere are rather weak \citep[see, e.g.,][]{DesMarais02, Vasquez13m, Schreier18ace} so it may be challenging to detect in exoplanetary atmospheres. In the Earth’s stratosphere, it is removed mainly via photolysis and reactions with electronically-excited oxygen atoms (O$^{*}$). Nitrous oxide can also be produced in the upper atmosphere via ionization of N$_2$ by solar activity \citep[see summary][]{Sinnhuber-etal-2012}, resulting in a possible “false positive” signal.
\\
\\
\textbf{Methane (CH$_4$) -} On Earth CH$_4$ has significant biological sources via methanogenic bacteria, although it also features somewhat weaker (by ~one order of magnitude) abiotic sources, for example, from outgassing. CH$_4$ is a key greenhouse gas and is mainly removed via oxidative degradation from Earth's atmosphere by the reactive hydroxyl (OH) radical. An overview of methane's global atmospheric budget inventory is provided by, e.g., \citet{Wahlen-1993}.
\\
\\
\textbf{Chloromethane (CH$_3$Cl) - } has atmospheric sources associated with, for example, seaweed emissions. It is removed, e.g., via photolysis and by the reactive OH radical \citep[see][]{Keppler-etal-2005}.
\section{(Exo)planetary Climate-Chemistry Modeling}\label{sec:2}
\label{sec:models}
This section deals with the incoming stellar radiation which interacts radiatively and photochemically as it passes through the planetary atmosphere. Numerical models (described below) are constructed to solve the central equations for radiative transfer, convection, and photochemistry. The key model outputs are vertical profiles of temperature, radiative fluxes, water (for climate-convection) and chemical abundances (for (photo)chemistry). 
\subsection{Earth-like planets in the HZ of M-stars}
1D modeling studies, for example by \citet{Segura-etal-2005}, \citet{Rauer-etal-2011}, and \citet{Rugheimer-etal-2015} applied a global mean, cloud-free coupled chemistry-climate column model to investigate biosignature responses for planets orbiting a range of M-dwarfs. Results suggested significant chemical responses, for example, a strong CH$_4$ greenhouse effect. However, these models did not include the effects of potentially high \ac{StEP} levels, which are expected to be essential for Earth-like planets orbiting M-dwarfs.

\citet{Segura-etal-2010} and most recently \citet{Tilley-etal-2017} used a proxy (based on X-ray observations, as described above) to calculate the NO$_x$ production from \ac{StEP} interaction with an Earth-like atmosphere (defined here as N$_2$-O$_2$ dominated), which they applied interactively in their global-mean coupled climate-chemistry column model and studied the effect on O$_3$ for Earth-like planets in the HZ of M-dwarfs. Their results suggest that O$_3$ may not survive strong \ac{StEP} events during flaring conditions due to catalytic losses by NO$_x$. 

\citet{Grenfell-etal-2012} also investigated the effects of \acp{StEP}, confirming that atmospheric O$_3$ may not survive extreme flaring cases for Earth-like planets orbiting active M-dwarfs. Furthermore, \citet{Tabataba-Vakili-2016} studied such effects with a similar model version but updated with an additional energetic particle-induced HO$_x$ source. In their work, a parameterization was introduced into the chemistry module simulating the breakup of molecular oxygen by SCRs which can then trigger ion cluster chemistry involving H$_2$O to ultimately produce HO$_x$, a sink for biosignatures such as O$_3$. \citet{Rauer-etal-2011} investigated theoretical spectral signals of Earth-like planets in the HZ of M-dwarfs (without considering energetic particles) for a range of stellar spectral classes and Earth-like planets with up to 3g, corresponding to about ten Earth masses. Results revealed important couplings between chemistry and climate which arose, e.g., middle atmosphere heating from enhanced planetary abundances of the greenhouse gas CH$_4$, especially for scenarios considering the cooler M-dwarfs that led to weakened biosignature spectral bands of, e.g., O$_3$. As a follow-up \citet{Grenfell-etal-2013} investigated the photochemical responses for the same scenarios as \citet{Rauer-etal-2011}. Results suggested that the mechanism for O$_3$ production switched from the Chapman mechanism to the smog mechanism for the cooler stars (which emit less (E)UV hence slow the Chapman mechanism since this proceeds via the photolysis of the strong O$_2$ molecule). Besides, \citet{Grenfell-etal-2013a} showed that varying the (uncertain) UV output of cool M-dwarfs could impact the photochemistry and have a potentially large effect on the O$_3$ spectral features. They also confirmed that N$_2$O spectral features could be enhanced for planets orbiting M-dwarfs. 

Presently, both 1D and 3D models are applied in the literature to investigate atmospheres of tidally-locked planets in the HZ of M-dwarfs. The two approaches are complementary since on the one hand 1D studies are useful to understand the wide and mostly unconstrained parameter range for exoplanet atmospheres, whereas 3D effects will play a central role because, e.g., tidally-locked planets could feature strong 3D transport cells from the day- to the nightside \citep[e.g. ][]{Joshi-2003}. More recently e.g., \citet{Joshi-Haberle-2012}, \citet{Leconte-etal-2013}, \citet{Shields-etal-2013}, \citet{Yang-etal-2013}, and \citet{Godolt-etal-2015} applied 3D models to study rocky exoplanets in the HZ of K- and M-dwarfs.

Moreover, the photochemical effects of \acp{GCR} upon biosignatures were studied by, e.g., \citet{Grenfell-etal-2007}. The results, however, suggest that the effects were quite modest for the scenarios investigated. For their study, they updated the standard column model chemistry module with a theoretical air shower approach to include the effects of energetic particles including related NO$_x$ sources. The results suggest that atmospheric biosignatures like O$_3$ and N$_2$O are mostly robust to \acp{GCR}, e.g., column O$_3$ decreased by only $\sim$20$\%$.
\subsection{Chemistry-climate effects in modern Earth's atmosphere}
At Earth, the impact of SCRs on the chemical composition of the atmosphere above 20 km, and particularly on nitrogen species and ozone, is well documented from satellite observations spanning back to the 1970s \citep[see, e.g.,][]{Crutzen-etal-1975,Swider-Keneshea-1973,Solomon-etal-1981, Solomon-etal-1983,Jackman-etal-1980,Jackman-etal-1990,Jackman-etal-2000,Jackman-etal-2005a,Jackman-etal-2005b,Rohen-etal-2005,Funke-etal-2011}.Ozone is the key species with respect to radiative heating in the terrestrial stratosphere. It is involved in important feedback, for example, coupling ozone heating with stratospheric temperatures and dynamical phenomena such as the Brewer-Dobson meridional circulation with its downward transport in the winter-spring polar 3D vortex on Earth can have a significant effect on the resulting O$_3$ loss \citep{Sinnhuber-etal-2003, Jackman-etal-2009}. Model studies using chemistry-transport models (CTMs) or chemistry-climate models (CCMs) including parameterizations for NO$_x$ and HO$_x$ production due to atmospheric ionization have been compared with simultaneous satellite observations from satellite instruments such as MIPAS on ENVISAT in a number of dedicated model-measurements intercomparison studies \citep[e.g.][]{Funke-etal-2011, Funke-etal-2017}. Key issues include the onset, distribution, and recovery time of the cosmic-ray induced photochemical ozone loss. Generally, a good agreement has been achieved in most studies between the observed and modeled NO$_x$ production and ozone loss during strong particle events \citep[e.g.][]{Funke-etal-2011}. However, recent observations and model studies also indicate a significant impact of complex ion-chemistry reactions during larger atmospheric ionization events \citep{Sinnhuber-etal-2012, Verronen-Lehmann-2013}. 
The impact of energetic particle ionization on chemical composition and dynamics (for the case of Earth) is summarized in \citet{Sinnhuber-etal-2012}.
\section{Simulation Models Utilized and Combined in this Study}\label{sec:AtRIS_1DTUB_ExoTIC}
\subsection{Particle Transport and Interaction Model: The \acf{AtRIS}}
AtRIS, the \textbf{At}mospheric \textbf{R}adiation \textbf{I}nteraction \textbf{S}imulator, was recently developed by \citet{banjac2018} to address the requirements necessary to investigate ionization in exoplanetary atmospheres. With AtRIS, the user can compute the hadronic and electromagnetic interactions of energetic particles with planetary atmospheres based on the GEneration ANd Tracking of particles version 4 (Geant4) code. A highlight feature is the customized \textbf{P}lanetary \textbf{S}pecification \textbf{F}ormat (PSF), which provides a simple yet extremely flexible interface for the specification of the exoplanetary soil and atmosphere. This code enables simulations of (amongst other things) (i) the atmospheric secondary particle flux, (ii) the energy loss (both in the atmosphere and in the ground) as well as (iii) the radiation dose on the surface of an arbitrary exoplanet over a wide variety of conditions.
\subsubsection{Short description of AtRIS}
There are ongoing efforts in the literature to improve understanding of both the electromagnetic radiation \citep[by, e.g., the MUSCLES and MEGA-Muscles project, see, e.g.,][, respectively]{France-etal-2016, Froning-etal-2018} and the flux of stellar super- and hyperevents \citep[see, e.g.,][]{Youngblood-etal-2017, Herbst-etal-2019a} experienced by rocky exoplanets orbiting M-dwarfs and K-stars. In the immediate future, we will not be able to measure precise primary TOA particle spectra for exoplanets. This was a starting assumption upon which the design of AtRIS has been based. Instead of simulating the effect that a specific primary particle spectrum has on the exoplanetary atmosphere, AtRIS calculates three so-called \textbf{A}tmospheric \textbf{R}esponse \textbf{M}atrices of the form $\left[M\right]_{m\times n}$, which quantify the relation between the energy (and species) of the primary particle and the ensuing altitude-dependent ionization, absorbed dose and dose equivalent. 
Thereby, the dose equivalent $H$ reflects the radiation hazard to the human body by taking into account radiation-type dependent biological effectiveness given by
$H=\sum_{i}W_{i}\cdot D_{i}$, where $W_i$ denotes the radiation weighting factor \citep[see][]{ICRP-2007} and $D_i$ the absorbed dose due to particles of type $i$. The $m$ rows of the matrices correspond to different altitudes (atmospheric shells), while the $n$ columns correspond to different primary energies. Each matrix element describes the average of the specific quantity \textit{per particle}. A column $\left[n\right]$ of the ionization ARM describes how a primary particle of such specific kinetic energy would, on average, deposit its energy at every single altitude throughout the atmosphere. Analogously, each matrix row $\left[m\right]$ describes how different primary energies contribute to the ionization at that specific altitude. 

By multiplying the matrix $M$ with a vector $\left[N\right], j\in\lbrace1,\dots,n\rbrace$, where $N_j$ describes the number of primary particles in the $j$\--th energy bin, we can quickly derive ionization, absorbed dose and dose equivalent for any given input spectrum. This approach has the advantage of being able to calculate the quantities of interest for different spectral shapes quickly. Furthermore, we can investigate the effects of a planetary magnetosphere and stellar modulation by imposing either a hard or a soft energy filter on the primary spectral vector $N$. 

AtRIS has been published under a general public license (GPL) and includes extensive online documentation in the form of a wiki page \citep{atrics_wiki}. This offers the unique opportunity for the exoplanetary community to make use of the above-described flexibility and unique features of AtRIS in order to achieve a better understanding of the exoplanetary atmospheric ionization and radiation environments. 
\subsubsection{Validation of AtRIS}
\citet{banjac2018} validated AtRIS for the following Earth-bound measurements: (i) ion-electron production rate \citep{neher1971cosmic}, (ii) surface neutron spectrum \citep{gordon2004measurement}, (iii) surface muon spectra \citep{kremer1999measurements,haino2004measurements}, (iv) balloon-borne proton spectrum \citep{haino2004measurements}, (v) absorbed and equivalent dose rates measured on a stratospheric-balloon \citep{moller-2013}, and (vi) absorbed dose rates measured during intercontinental flights \citep{moller-2013}. Further studies for Mars and Venus have been published by \citet{Guo-etal-2019} and \citet{Herbst-etal-2019b}, respectively.
\subsection{Magnetospheric transport}
Since AtRIS does not implement the geomagnetic field, we used PLANETOCOSMICS \citep{desorgher2006planetocosmics}, which has been previously applied to study the effect of geomagnetic field geometry upon particle propagation to the Earth’s surface \citep[see, e.g.,][]{Fichtner-etal-2012, Herbst-etal-2013}. Thereby, the following magnetic fields are implemented: the International Geomagnetic Reference Field (IGRF), various magnetic dipole fields, Mars-like crustal fields as well as more advanced models taking into account the interaction with the (possibly strong) stellar wind, like the model by \citep{tsyganenko1996modeling} (TSY96), which uses the stellar wind speed and magnetic field strength as input parameters. Based on these, we simulated the location and altitude-dependent cutoff rigidity values, which are a measure of the energy a particle must have in order to reach a specific point in the planetary atmosphere. The TOA proton flux is calculated depending on the cutoff rigidity.  
\subsection{Ion-neutral chemistry Model}

The \textbf{Exo}planetary \textbf{T}errestrial \textbf{I}on \textbf{C}hemistry model (ExoTIC) is an adapted version of the \textbf{U}niversity of \textbf{B}remen \textbf{I}on \textbf{C}hemistry column model (UBIC). UBIC is a state-of-the-art one-dimensional stacked box model of the neutral and ion atmospheric composition originally developed in the group of M. Sinnhuber at the University of Bremen to investigate the impact of energetic particle precipitation on ion and neutral chemistry in the terrestrial D- and E- region \citep{Winkler-etal-2009, Sinnhuber-etal-2012, Nieder-etal-2014}. It is now maintained and further developed at the \ac{KIT}. Neutral and ion photochemistry is driven by photo-absorption in the UV, visible and near-IR range (120-800~nm) as well as by photo-ionization in the EUV range and by particle impact ionization. Photolysis and photoionization rates are calculated from the stellar spectrum provided, particle impact ionization rates need to be prescribed. The primary charge due to the atmospheric ionization is distributed amongst primary ions depending on the energy range of the ionization sources as well as on the atmospheric composition. UBIC considers 60 neutral and 120 charged species, which interact due to neutral, neutral-ion, and ion-ion gas-phase reactions, as well as photolysis and photo-electron attachment and detachment reactions \citep[see, e.g.,][]{Sinnhuber-etal-2012}. Temperature and pressure as a function of altitude also need to be prescribed; the vertical coverage of the model depends on the availability of temperature/pressure profiles. ExoTIC extends UBIC's applicability to atmospheres of terrestrial (rocky) planets other than Earth. 

Within our model suite, particle impact ionization rates will be prescribed by output from the \ac{AtRIS} code as a function of the activity of the star. Temperature and pressure profiles are provided by the output from the Coupled 1D Terrestrial Climate-Chemistry Model (1D-TUB, see below). 

In our study, the effects of ion chemistry from the ExoTIC model are implemented into the 1D-TUB model as follows: production and loss rates due to ion chemistry from ExoTIC affecting a range of neutral species are implemented into the 1D-TUB model \citep[see also][]{Nieder-etal-2014}. The ExoTIC rates are calculated specifically for the composition and temperature-pressure profiles output by the 1D-TUB model for a given iteration and then provided as input for that model for the next iteration (see discussion Sec. 5). In this way, atmospheric profiles for rates of change due to ion chemistry for the following ten species relevant for HOx and NOx responses: NO, NO$_2$, NO$_3$, HNO$_3$, N$^4$S, N$^2$D, H, OH, O and O$^1$D were implemented from ExoTIC into the 1D-TUB model. Future work will consider additional species.
\subsection{Coupled 1D Terrestrial Climate-Chemistry Model (1D-TUB)}\label{cccc}
We use a 1D, steady-state, cloud-free radiative-convective module to calculate atmospheric temperatures and water content, coupled to a photochemistry module calculating the steady-state mixing ratios of 55 species with over 200 reactions. This model, dating back to \citet{Kasting-etal-1984} , \citet{Pavlov-etal-2000}, and \cite{Segura-etal-2003}, has been subsequently developed within the DLR/TUB Berlin group \citep[][]{Grenfell-etal-2011, Rauer-etal-2011, vonParis-etal-2015, Tabataba-Vakili-2016, Gebauer-etal-2017, Gebauer-etal-2018a, Scheucher18}.

Please note that we do not explicitly include ion chemistry in the 1D-TUB climate-chemistry column model. Thus, we do not include ion reactions in our chemical network since this model was initially developed for the lower layers of Earth's atmosphere (in particular the mesosphere and below) where ion chemistry is not typically significant. The effects of ion chemistry, however, were parameterized in our model by using input from the partner model ExoTIC (see description above).
\subsubsection{Climate Module}\label{sec:climate} 
The climate module of 1D-TUB discretizes the planetary atmosphere into 101 layers equally spaced in log(p) from the surface up to 1 Pa. This region was chosen in order to simulate characteristic radiative photochemical processes in Earth's neutral (lower to the middle) atmosphere ranging from the troposphere and stratosphere up to the mesosphere. The radiative-convective scheme is adapted from \citet{Toon-etal-1989} for the shortwave stellar absorption and Rayleigh scattering. The Modified RRTM for Application in CO$_2$-dominated atmospheres \citep[MRAC,][]{vonParis-etal-2015} using the correlated-k approach, is used for thermal molecular absorption by H$_2$O, CO$_2$, CH$_4$, and O$_3$. For the lower atmosphere, convective adjustment to a wet adiabatic profile is performed, and the water profile is parameterized after \citet{Manabe-1967}. The model calculates 1D column global-average, cloud-free conditions, although cloud effects are considered straightforwardly by adjusting the surface albedo until the mean surface temperature of the Earth (288.15K) is achieved for the Earth control scenario.
\subsubsection{Chemistry Module}\label{sec:chemistry} 
The photochemistry module was adapted from \citet{Segura-etal-2003} for various planetary conditions. The chemistry module of 1D-TUB calculates the atmospheric conditions on the same pressure layers from the climate module minus the surface layer, hence 100 layers in total.  \citet{Gebauer-etal-2017} introduced variable O$_2$ content for early Earth's atmosphere, while \citet{Gebauer-etal-2018a} updated the O$_2$ photolysis rate. For the photochemistry effects induced by the precipitating high energy particles (GCRs, SEPs) we have two possibilities, namely i) an air shower approach using the Gaisser-Hillas method, as detailed in \citet{Tabataba-Vakili-2016} and further developed as described in \citet{Scheucher18}, and ii) we can directly process the cosmic-ray-induced ionization rate of the atmosphere as calculated by AtRIS and the cosmic-ray-induced chemical production rates of atmospheric molecules calculated by ExoTIC. In both approaches, the photochemistry module then incorporates atmospheric profiles of these elaborate cosmic-ray-induced changes in composition. The chemistry routine also includes biogenic and source gas emissions such that Earth’s modern-day surface abundances can be reproduced.
\begin{center}
\begin{figure*}[!t]
 \includegraphics[width=\textwidth]{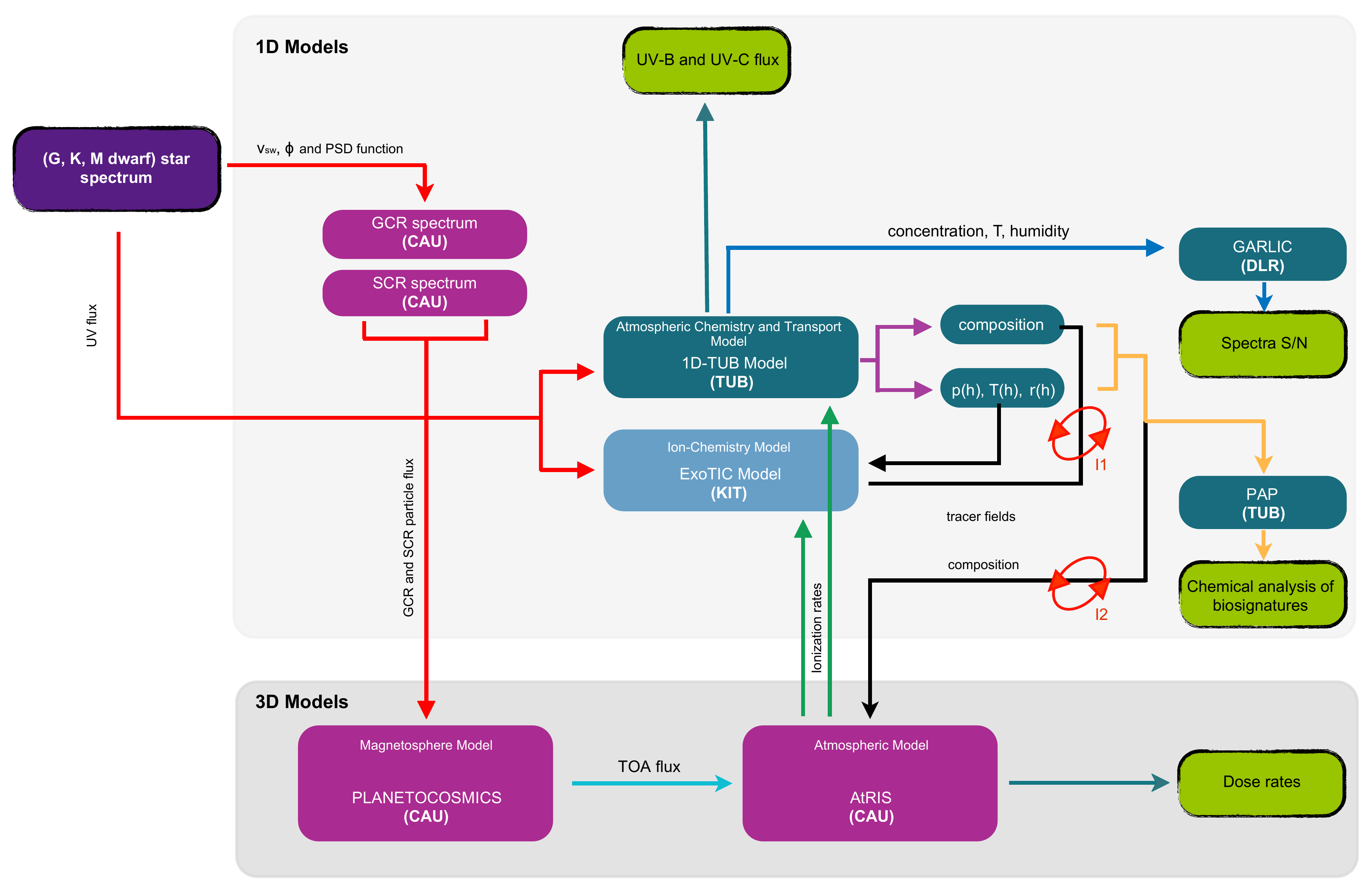}
\caption{Data exchange in the model suite and associated scientific goals. The following steps are performed: \textcolor{red}{(1)} (top left) measured stellar UV- fluxes as well as the estimates of the stellar wind velocities ($v_{sw}$), stellar modulation ($\phi$) and PSDs (Christian-Albrechts-Universit\"at, hereafter CAU) are used as input i) for the Berlin Coupled Column Climate Chemistry (1D-TUB) and ExoTIC models as well as ii) to estimate the incoming GCR and SCR fluxes at the close-in exoplanets, respectively (highlighted by the red paths) \textcolor{cyan}{(2)} the transport of CRs through the planetary magnetosphere is studied with PLANETOCOSMICS in order to provide TOA proton fluxes as input to model both the GCR and SCR induced atmospheric ionization as well as the resulting dose rates on the planetary surface (purple arrow), (3) computation of the secondary particle production due to CR interaction through the planetary atmosphere down to the surface, \textcolor{Skobeloff}{(4)} calculation of the surface UV-A, UV-B and UV-C exposure as well as the radiation dose, \textcolor{ForestGreen}{(5)} determination of the impact of changing atmospheric ionization for the different atmospheric compositions as well as parameterization of the neutral atmosphere impact, \textcolor{Mulberry}{(6)} computation of the resulting atmospheric composition and climate, \textcolor{Dandelion}{(7)} performance of a pathway analysis in order to understand the biosignature chemical responses, \textcolor{RoyalBlue}{(8)} utilization of the global atmospheric composition and temperature fields to compute atmospheric transit (primary) and emission (secondary) spectra. The two iterations within the framework are displayed as red circles.}
 \label{fig:3}
 \end{figure*}
\end{center}
\subsection{Diagnostic Tools}
\subsubsection{The Pathway Analysis Program}
The Pathway Analysis Program, PAP, identifies and quantifies chemical pathways that produce or destroy a chemical species of interest. PAP takes as input species concentrations and reaction rates 
provided by a chemical model. Operating step-by-step (i.e., one species after another), PAP links reactions which form a given chemical species with reactions which destroy it. By specifying O$_3$ as the species of interest, we can apply PAP to identify and quantify O$_3$-producing pathways as well as catalytic cycles that destroy O$_3$. PAP also determines the corresponding rates, which allow us to assess the importance of individual cycles at different locations (altitudes). For instance, catalytic cycles involving nitrogen oxides, chlorine oxides, and hydrogen oxides are well-known to destroy O$_3$ from the lower to the upper stratosphere of the Earth \citep[e.g.,][]{Jucks-etal-1996}. PAP can however also be applied to conditions which differ significantly from those on Earth. It is, therefore, an invaluable tool for understanding complex O$_3$ chemical responses in exoplanetary atmospheres. The PAP was developed by \citet{Lehmann-2004} and applied by \citet{Grenfell-etal-2007} to Earth’s stratosphere, by \citet{Stock-etal-2012} to the Martian atmosphere, by \citet{Grenfell-etal-2013} to Earth-like exoplanets in the HZ around M-stars, by \citet{Gebauer-etal-2017} and \citet{Gebauer-etal-2018a} for early-Earth and early Earth analog planets, and by \citet{Gebauer-etal-2018b} for early Earth analogs around M-stars.
\subsubsection{The Generic Atmospheric Radiation Line-by-Line Infrared Code}
The Generic Atmospheric Radiation Line-by-line Infrared Code or GARLIC \citep{Schreier14} has originally been developed for Earth atmospheric remote sensing and, more recently, has been used extensively for exoplanet studies \citep[e.g.][]{Rauer-etal-2011, Scheucher18}. The core of GARLIC's subroutines constitutes the basis of forward models used to implement inversion codes to retrieve atmospheric state parameters from limb and nadir sounding instruments.
GARLIC has been verified in several intercomparison studies \citep[e.g.][]{Schreier18agk} and validated against spectra observed on Earth \citep{Schreier18ace} and from Venus \citep{Hedelt13,Vasquez13v}. GARLIC utilizes molecular line data from the HITRAN or GEISA database \citep{Gordon17etal, JacquinetHusson16etal} with optional corrections for continuum and collision-induced absorption and calculates theoretical atmospheric spectra (emission, transmission, effective height) based on the converged output (climate and abundance profiles) from 1D-TUB.
\section{Building up a self-consistent simulation chain}
Figure~\ref{fig:3} sketches the interplay between the modules of our coupled model suite and the goals that we will achieve. A close interaction between the three groups involved (Karlsruher Institute of Technology [KIT], Technical University Berlin [TUB], and CAU) is mandatory. Additional interfaces to address atmospheric climate and composition as well as the assessment and parameterization of ion-chemical impacts between models have been developed. The following steps have to be performed: (1) The measured stellar UV- fluxes, as well as the estimates of the stellar wind velocities, stellar modulation and PSDs (CAU), will be used as input i) for the 1D-TUB and ExoTIC model as well as ii) to estimate the incoming GCR and SCR fluxes at the close-in exoplanets, respectively. (2) With the latter, the transport of cosmic rays (CRs) through the planetary magnetosphere is studied with PLANETOCOSMICS in order to provide TOA proton fluxes as input to model both the GCR and SCR induced atmospheric ionization as well as the resulting dose rates on the planetary surface. (3) Computation of the transport of CRs through the planetary magnetosphere to provide TOA proton fluxes. (4) Computation of the secondary particle production due to CR interaction through the planetary atmosphere down to the surface. (5) Calculation of the surface UV-A, UV-B, UV-C exposure, and the radiation dose. (6) Determination of the impact of changing atmospheric ionization for the different atmospheric compositions as well as parameterization of the neutral atmosphere impact. (7) Computation of the resulting atmospheric composition and climate and especially the atmospheric neutral and ion chemistry of biosignatures (1D) for the different scenarios. (8) Performance of a pathway analysis (varying in season and location) in order to understand the biosignature chemical responses. (9) Utilization of the global atmospheric composition and temperature fields to compute atmospheric transit (primary) and emission (secondary) spectra. (10) Validation of the 1D scenario between the different models for Earth-like conditions. Note that two iterations within the framework are performed (red circles). Since the neutral species modeled with ExoTIC (KIT) strongly depend on the ionization rates provided by CAU,  iterations have to be performed (I1). After the first run, the computed neutral species are handed to TUB in order to determine the corresponding atmospheric composition, which is then provided as new input for AtRIS (I2). These two iterations are performed until no changes in neutral components and atmospheric composition due to the computed ionization rates occur.
%
%
\begin{figure}[!t]
\begin{center}
  \includegraphics[width=0.7\columnwidth]{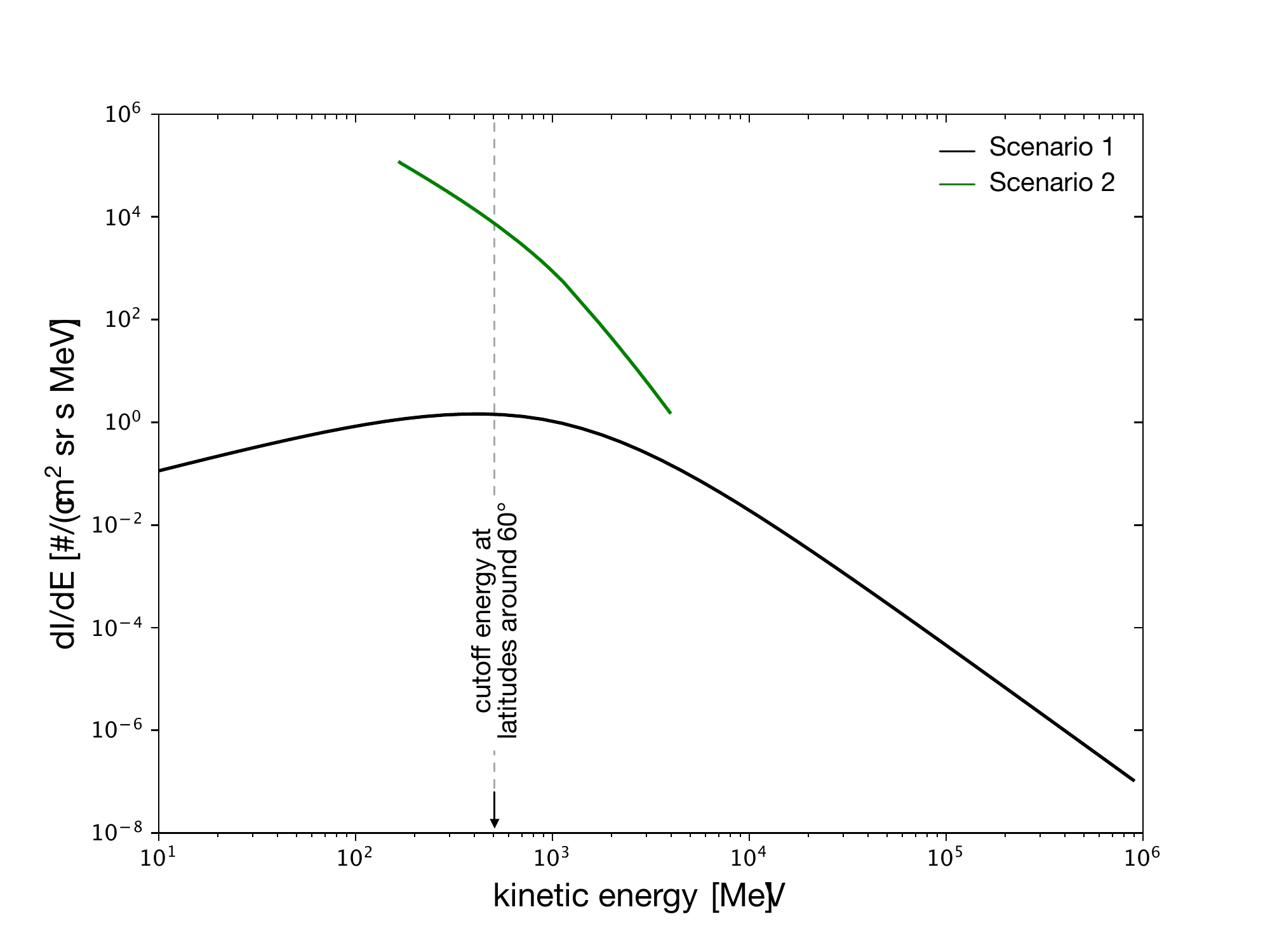}
  \caption{Differential primary proton GCR spectrum during solar minimum (quiescent sun) conditions (scenario 1, black line) in comparison to the strongest ever measured Ground Level Enhancement (GLE) event (GLE05, February 23, 1956) (scenario 2, green line). The gray dashed line indicates the cutoff energy, i.e., the energy a particle must have in order to reach a certain location within the terrestrial atmosphere at mid-latitudes around 60$^\circ$.}
 \label{fig:4}
 \end{center}
 \end{figure}
\subsection{Testing the model suite on the (exo)planetary case-study of the modern-day Earth orbiting a G2V star}
This section studies the impact of GCRs and SCRs on the terrestrial atmosphere as an initial validation for our model-suite.
\subsubsection{The GCR and SCR environment at 1~AU}\label{scenarios}
As discussed in Sec.~2.1 the GCR flux at 1~AU strongly depends on the solar activity and can be described by the \ac{FFA} (see Eq.~(\ref{Eq.2})). Thereby, the only free parameter, the solar modulation parameter $\phi$, is directly proportional to the solar activity. PAMELA and AMS02 measurements suggest that typical modulation parameter values vary between 500 and 2000 MV. During quiet solar conditions, where the modulation effect is less pronounced, more low energetic GCR particles can be observed in the Earth's vicinity. Thus, solar minimum conditions are in favor to be used in order to investigate the upper limit of the GCR-induced atmospheric ionization. The quiescent GCR flux based on the LIS model by \citet{Herbst-etal-2017} is displayed as solid black line in Fig.~\ref{fig:4}. 

In addition, the mean energy spectrum of the most strongest GLE event ever measured during the Neutron Monitor era, GLE05 that occurred on February 23, 1956, is given as solid green line. Note that the spectrum is based on the Band-Fit function proposed by \citet{Raukunen-etal-2018}. Representing one of the strongest GLEs in solar history this spectrum provides a good approximation for an upper limit of SEP flux. However, two things become apparent in Fig.~\ref{fig:4}: (i) solar particles with energies well above 1~GeV are rare, and (ii) the differential proton intensities measured during such GLEs can be in the order of one to more than four orders of magnitude higher than the quiescent GCR flux.
\subsubsection{The GCR- and SCR- induced atmospheric ionization, ion chemistry, and climate}
Cosmic rays penetrating the Earth’s atmosphere mainly lose their energy due to ionization of ambient matter. Thereby, the probability of an interaction with the surrounding atmospheric gases drastically increases the further the induced nucleonic-electromagnetic particle shower evolves towards the surface. Thus, the production of charged secondary particles also leads to the ionization of the middle as well as the lower atmosphere. This, however, strongly depends on i) the energy of the primary particle, ii) the secondary particle type as well as the iii) the atmospheric depth \citep[see, e.g.,][]{Bazilevskaya-etal-2008, Usoskin-etal-2011, banjac2018}.

Since GCRs are modulated due to the solar activity the corresponding ionization is anti-correlated to the eleven-year solar cycle. In case of SEP events, however, only very high energetic SCRs, such as the particles accelerated in strong SEP events, can induce and contribute significantly to the atmospheric ionization.

Numerically speaking, the cosmic ray induced ionization rate $Q$ is given by:
\begin{equation}
    Q(\phi, E_C, x) = \sum_i \int_{E_{C, i}}^{\infty} J_i(\phi, E) \cdot Y_i(E,x)~dE,
\end{equation}
depending on solar activity, geomagnetic location (here in form of the cutoff energy $E_C$ of a particle necessary in order to reach a certain location), and atmospheric depth ($x$). Thereby, $i$ is the cosmic ray particle type, $J_i(\phi, E)$ the primary differential energy GCR or GLE spectrum and $Y_i(E,x)$ the atmospheric ionisation yield given by $Y_i(E,x)$ = $\alpha \cdot \frac{1}{E_{ion}} \cdot \frac{\Delta E_i}{\Delta x}$, depending on the geometrical normalization factor $\alpha$ given by 2$\pi \int \cos(\theta) \sin{\theta}~d\theta$, the depth-dependent mean specific energy loss as well as the atmospheric ionization energy $E_{ion}$. For the latter most often an average ionization energy of 35 eV \citep[see][]{Porter-etal-1976} is used. Note, however, that a study by \citet{Wedlund-etal-2011} showed that this value strongly depends on the atmospheric composition, and that for modern Earth conditions a value of 32$\pm$1 eV is more likely. This will be of great importance for the ionization of exoplanetary atmospheres with differing atmospheric compositions \citep[see, e.g., discussion in][]{Herbst-etal-2019b}.

In the following studies, we assumed mid-latitudes around 60$^\circ$. At such locations on Earth, the corresponding cutoff energy is in the order of 510 MeV (indicated by the dashed vertical line in Fig.~\ref{fig:4}). Particles with energies below this threshold are not able to penetrate the geomagnetic field and cannot, therefore, contribute to the atmospheric ionization. The computed ionization rates have a strong influence on the computation of atmospheric neutral species hence on the atmospheric composition (see discussion below). Ionization rates are computed utilizing the initial climate and composition data from the 1D-TUB model ("Modern Earth Atmosphere", TUB). These rates are then implemented in both the 1D-TUB and ExoTIC model to provide an event-based set of atmospheric neutral species and atmospheric compositions.

As discussed in Sec. 5, it is crucial to reach stable neutral component and atmospheric composition conditions. In order to achieve those, multiple iteration rounds between the different models (1D-TUB, AtRIS, and ExoTIC) may be necessary.
\\
\\
In this study, we investigate two different scenarios: 
\begin{itemize}
    \item the influence of the quiescent Sun ($\phi = 550$ MV) only taking into account GCRs, hereafter \textit{scenario 1}
    \item the influence of the Febrary 23, 1956 ground level enhancement (GLE05), one of the strongest GLEs ever measured, hereafter \textit{scenario 2} 
\end{itemize}
%
%

\begin{figure}[!t]
\begin{center}
\includegraphics[width=0.7\columnwidth]{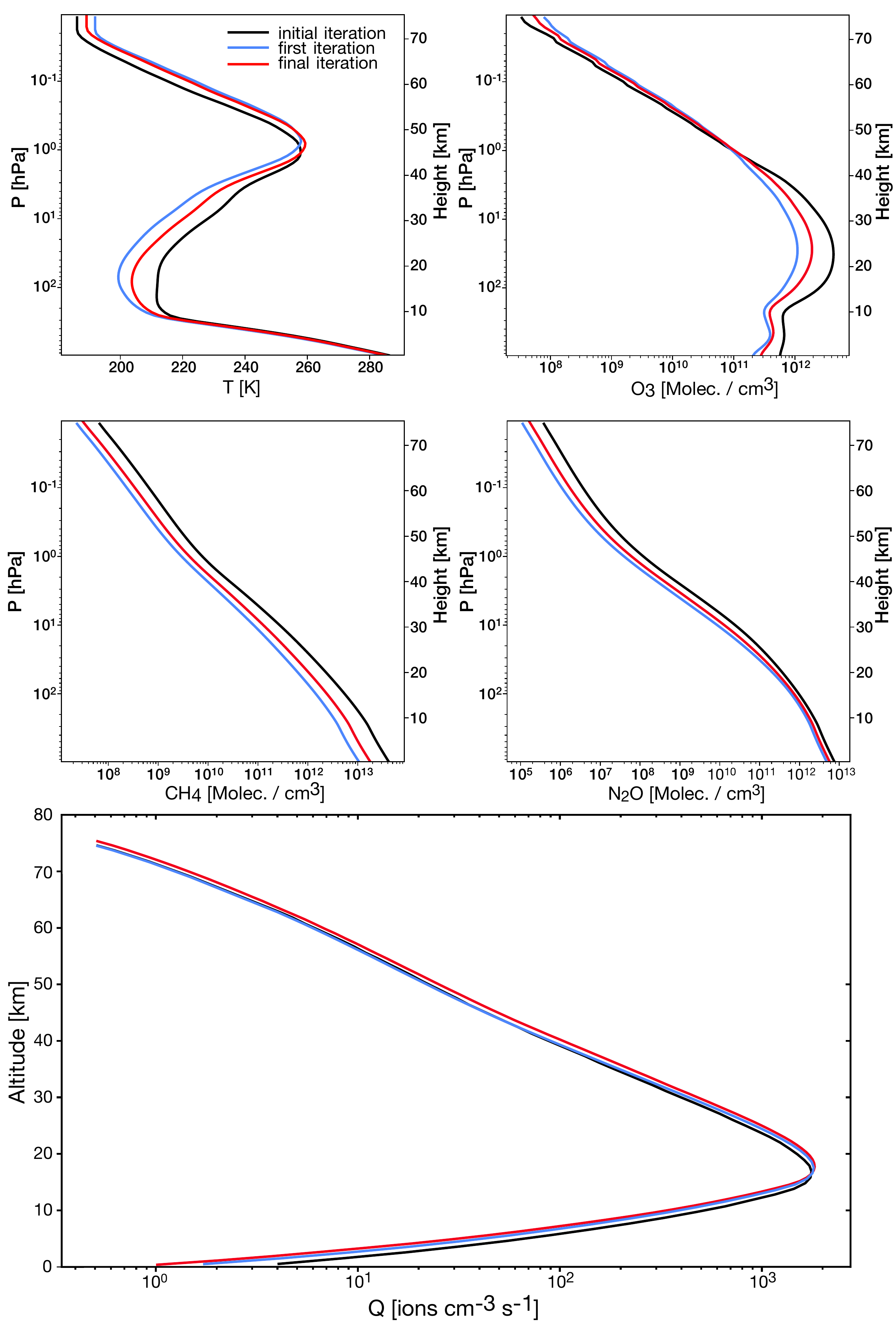}
  \caption{Model results for scenario 2 (GLE05). Three simulation rounds were mandatory in order to achieve converged output. Here, the initial conditions are displayed in black, while the first and second iteration are shown in blue and red, respectively. Note that the results of the second iteration have been used in a third iteration. Since the results converged the third iteration are similar to those of the second and are therefore not shown here. Note further, that the initial conditions (solid black line) are the same as the final conditions of scenario 1.}
 \label{fig:5}
 \end{center}
\end{figure}

%
In the case of scenario 1 only one iteration between the three models was necessary, whereas for scenario 2 more iterations were mandatory. The different iteration-dependent results of scenario 2 are shown in Figure~\ref{fig:5}. Here the corresponding temperature profile (upper left panel), as well as the molecular profiles of O$_3$ (upper right panel), CH$_4$ (middle left panel) and N$_2$O (middle right panel) provided via the 1D-TUB model, are given. As mentioned previously, the atmospheric compositions of the initial run (solid black lines) are based on scenario 1. Implemented into AtRIS and induced to a much harsher SCR environment the resulting ionization rate profiles (see bottom panel) that are provided as input for the ExoTIC model will lead to changes of the initial atmospheric profiles of the 1D-TUB model. In total, three iterations between the models were mandatory in order to achieve a stable neutral component and atmospheric composition conditions. It shows that the initial conditions (solid black line) strongly differ from the final iteration conditions (red lines). Moreover, comparing the final temperature profiles that reproduce Earth's characteristic temperature-layered regions (upper left panel) for scenario 1 (black solid line) with those of scenario 2 suggests that the profiles are similar at altitudes below $\sim$ 10 km, while differences of up to 10 K can be observed at higher altitudes. Thereby, the temperatures are lower up to an altitude of $\sim$ 45 km, while being slightly higher at altitudes above due to increased O$_3$ abundances (see also Fig.~\ref{fig:8}).

In addition, the climate model discussed in Section~\ref{sec:climate} featured an input surface albedo value, which was adjusted step-by-step until the observed global mean temperature for modern Earth equal to 288 K was reached. This is a rather common procedure in the Earth-like literature \citep{Segura-etal-2003} and is equivalent to placing an artificial reflecting "cloud" layer at the surface which is intended to compensate straightforwardly for the lack of cloud microphysics in the model. The surface albedo value in the model (= 0.232) is accordingly somewhat higher than the global mean observed value of ~0.15 \citep{Hummel-Reck-1979}. The convective and radiative modules could successfully reproduce the main features of the modern Earth's global mean temperature profile including the tropospheric lapse rate, the magnitude of the cold trap, and
the temperature inversion in the middle atmosphere due to heating from ozone absorption.

The chemical network discussed in Section~\ref{sec:chemistry} calculates a stationary, global-mean ozone column of 313.5 Dobson Units for modern Earth conditions. This compares well with observations in the modern Earth's atmosphere \citep{WMO-1994}. Ozone
features rather complex photochemical-climate responses, so the reproduction of its global column is a good performance test of the
chemistry module. Long-lived source gases such as carbon dioxide (CO$_2$) were set to 1990 levels at the surface (= 355 ppm) consistent
with previous model studies of Earth-like atmospheres \citep[see, e.g.,][]{Segura-etal-2003}. Surface concentrations of biomass and related gases such as methane (CH$_4$), chloromethane (CH$_3$Cl),  nitrous oxide (N$_2$O) and carbon monoxide (CO) were also fixed to modern values. The corresponding global mean mass fluxes required to maintain these surface concentrations compared reasonably well with modern Earth
observations \citep[see, e.g.,][for more details on the procedure]{Grenfell-etal-2014}.
%

\begin{figure}[!t]
\begin{center}
  \includegraphics[width=0.7\columnwidth]{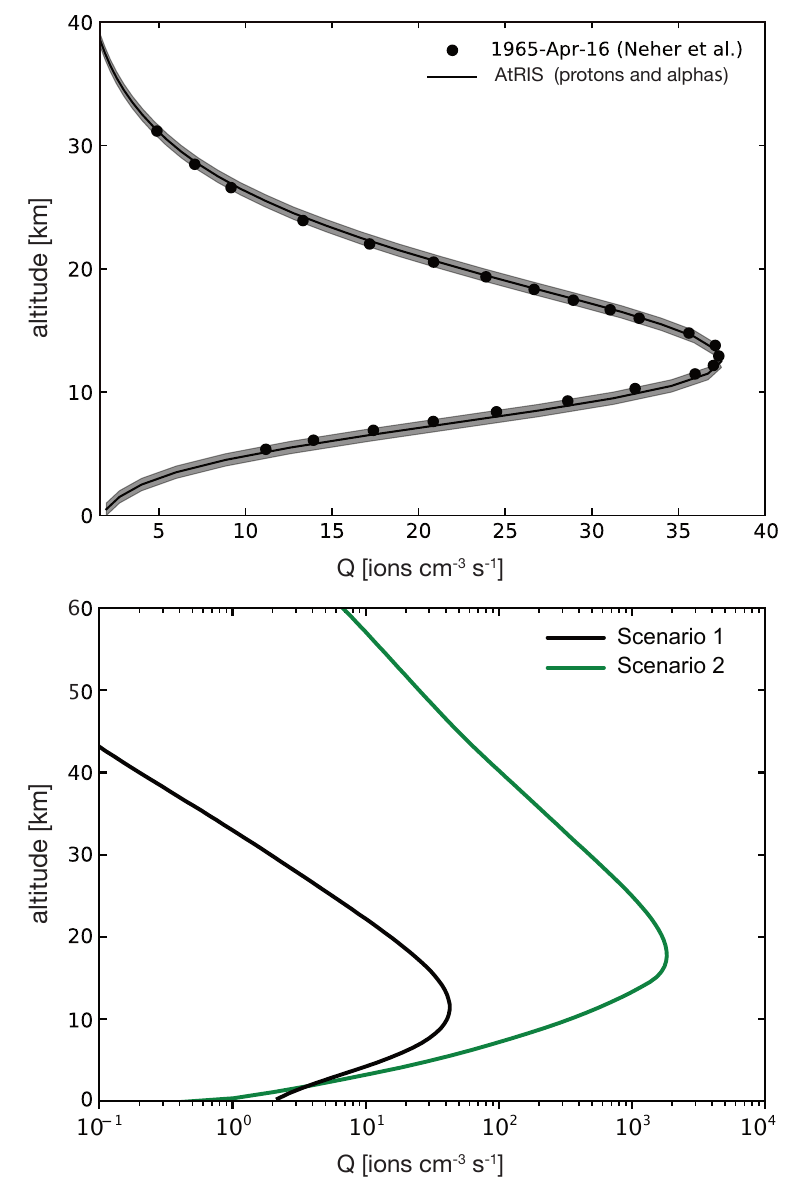}
  \caption{Upper panel: Validation of the modeled GCR-induced ion-pair production rates (solid line) against balloon measurements performed by \citet{neher1971cosmic} (dots) \citep[see][for further validation]{banjac2018}. Lower panel: CR-induced atmospheric ionization for scenario 1 (quiescent Sun) and scenario 2 (GLE05) displayed as black and green curves respectively. As can be seen, differences of up to three order of magnitudes occur.}
 \label{fig:6}
 \end{center}
 \end{figure}

\begin{figure}[!t]
\begin{center}
\includegraphics[width=0.7\columnwidth]{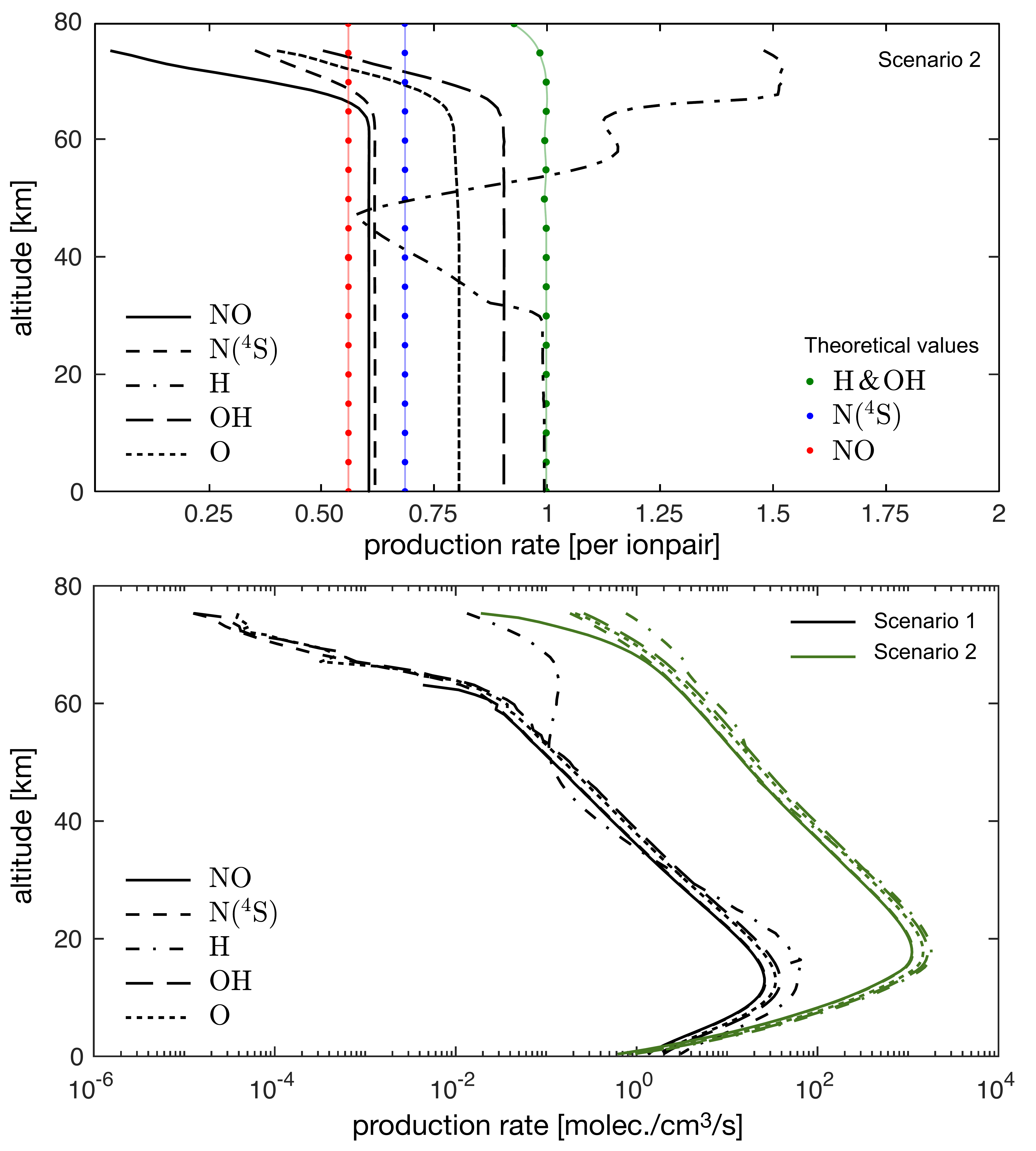}
  \caption{Upper panel: Production rates of neutral species NO, N($^4$S), H, OH and O due to ion-chemistry processes calculated with ExoTIC (black lines) for scenario 2 (GLE05) compared to theoretical values of existing parameterizations (coloured dots, see text). Lower panel: Production rates of NO, N($^4$S), H, OH and O due to ion-chemistry processes in the atmosphere for scenario 1 (quiescent Sun, black lines) and scenario 2 (GLE05, green lines).}
 \label{fig:chemistry}
 \end{center}
 \end{figure}

%
Figure \ref{fig:6} shows our modeled ionization rate profiles. Here, the upper panel shows a comparison between the altitude-dependent ionization profile measured during a balloon flight on April 16, 1965 \citep[black dots][]{neher1971cosmic} and the model results of AtRIS (black line). This Figure suggests good agreement between simulations and the measurement. A comparison of the scenario-based stable-condition ionization rate profiles is given in the lower panel Fig.~\ref{fig:6}. As can be seen, the strong solar energetic particle event of scenario 2 (green curve) leads to drastically increased altitude-dependent ionization rates at all altitudes compared to the results of scenario 1 (black curve). Differences of up to three orders of magnitude can be found. These are also reflected in the production rates of neutral species, as can be seen in Fig.~\ref{fig:chemistry}.

As described in section 4.3, thereby, ion chemistry processes influence the concentrations of the neutral constituents of the atmosphere. For the two scenarios discussed here, this mainly affects NO, N($^4$S), H, OH, and H$_2$O. The upper panel of Fig.~\ref{fig:chemistry} shows a comparison of the production rates derived with the full ion chemistry calculation using ExoTIC with simple parameterization models as summarized in \cite{Jackman-etal-2005a}. The parameterizations consider the terrestrial homosphere with a fixed mixing ratio of N$_2$ and  O$_2$. The parameterization for N($^4$S) and NO is originally based on \cite{Porter-etal-1976}. It only considers the direct impact of dissociation and dissociative ionization, no ion chemistry. This leads to a strong deviation above $\approx$70~km, where the impact of ion chemistry becomes more important and where the primary ions vary due to the continuing photolysis of O$_2$, see, e.g., \cite{Nieder-etal-2014}. Below this altitude, production rates of N($^4$S) and NO are constant functions of the ion pair production rate, though the factors differ because ExoTIC also considers the formation of NO by ion chemistry, leading to a higher formation rate of NO than in the parameterization. The parameterization of H and OH is based on \cite{Solomon-etal-1981} and considers a simplified set of protonized cluster ion reactions for the homosphere and a fixed water vapor profile. It also deviates strongly from the ExoTIC values above 70~km altitude presumably because different water vapor profiles are used, and due to the more complex ion chemistry considered in ExoTIC. Below this altitude, the rate of OH production is about 10~\% lower in ExoTIC than in the parameterization. This has already been discussed in \cite{Sinnhuber-etal-2012}, and is due to the formation of protonized water cluster ions via $NO^+(H_2O)_3 + H_2O$, which do not release OH. In the altitude range 40--60~km, the production rate of H as calculated by ExoTIC varies with altitude, deviating strongly from the parameterization values. In this altitude region, a transition occurs from recombination of the protonized water cluster ions by negative ions to recombination by electrons (not shown). Apparently, this transition is not smooth, leading to the observed dip in the H production. These differences between results from ExoTIC considering full ion chemistry to parameterizations considering the homosphere (either with or without simplified, ion chemistry) emphasize the importance of ion chemistry; for exoplanets with atmospheric compositions widely different to the terrestrial atmosphere, it is mandatory to explicitly consider the formation of primary ions and the subsequent effects of ion chemistry upon the neutral constituents.  The lower panel of Fig.~\ref{fig:chemistry} displays the modeled production rates of these species due to ion-chemistry processes in the atmosphere for the final iteration based on the inputs from the 1D-TUB model and the ionization rates provided by AtRIS. Here, the production rates for scenario 1 are displayed in black, those of scenario 2 in green.
%

%
\begin{figure*}[!t]
 \includegraphics[width=\textwidth]{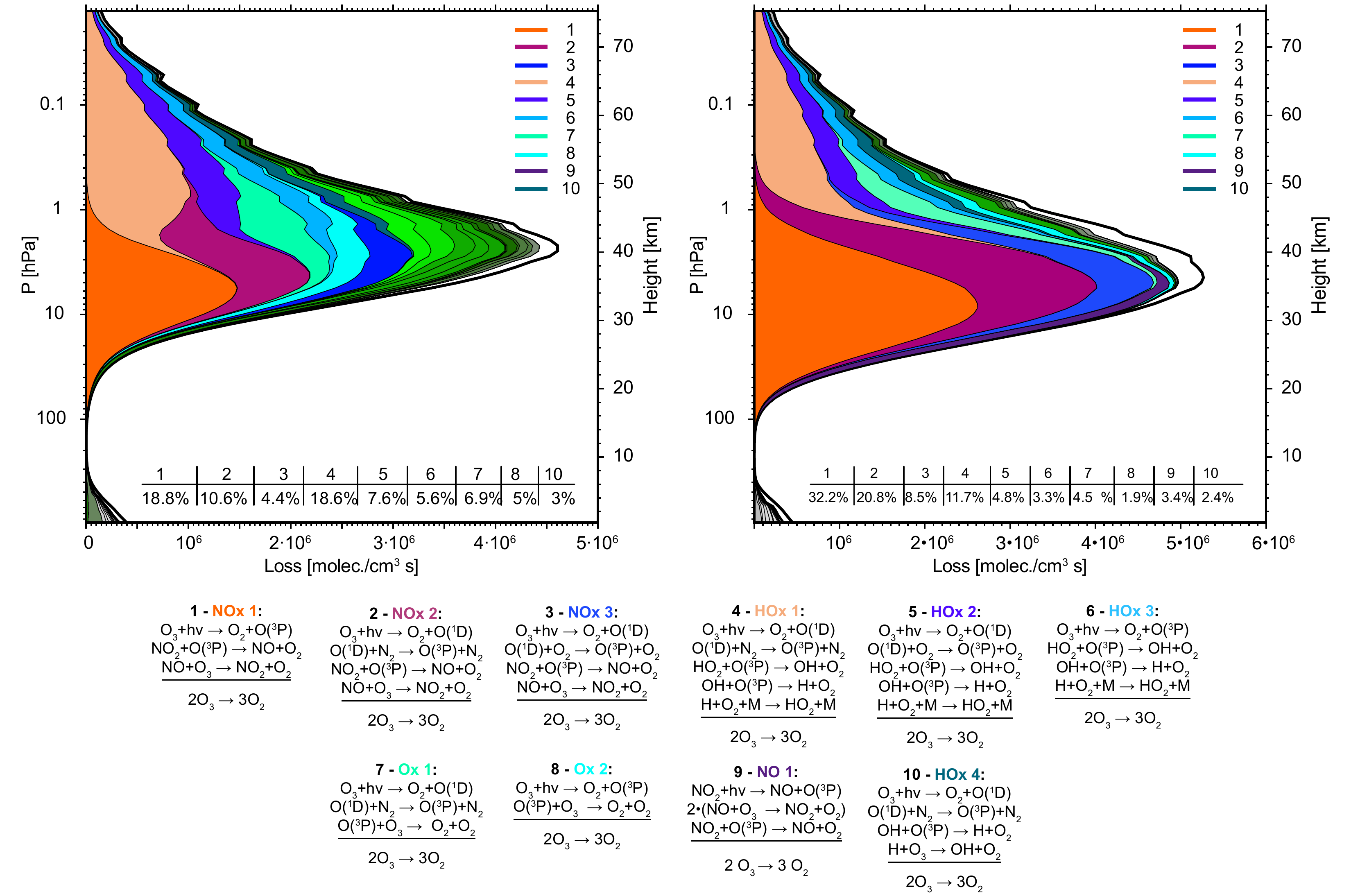}
  \caption{Ozone loss cycles determined by PAP corresponding to scenario 1 (quiescent Sun, left panel) and scenario 2 (GLE05, right panel) are shown on a linear scale from the surface to the mid mesosphere ($\sim$ 75km). Please note that the figures show all pathways with a percentage contribution above 0.01$\%$ and that only the ten most important pathways are tabulated in the subpanel.}
 \label{fig:11}
 \end{figure*}
\section{Impact of cosmic rays on biosignatures as well as the surface UV-radiation and radiation dose}
Since the CR-induced atmospheric ion-pairs have a direct influence on the atmospheric radiation dose, the atmospheric chemistry and, thus, on potential biosignatures like O$_3$, N$_2$O or CH$_4$, the results of Fig.~\ref{fig:6} as discussed in the following section are used as input to model the influence on the planetary atmosphere, addressing crucial questions like: How do CRs and the resulting photochemistry affect
the abundance of potential biosignatures? What is the radiation dose at the surface of Earth and Earth-like planets and how strong is the surface UV radiation exposure? How do CRs affect the spectral features of
biosignatures?
%
\begin{figure*}[!t]
 \includegraphics[width=0.49\textwidth]{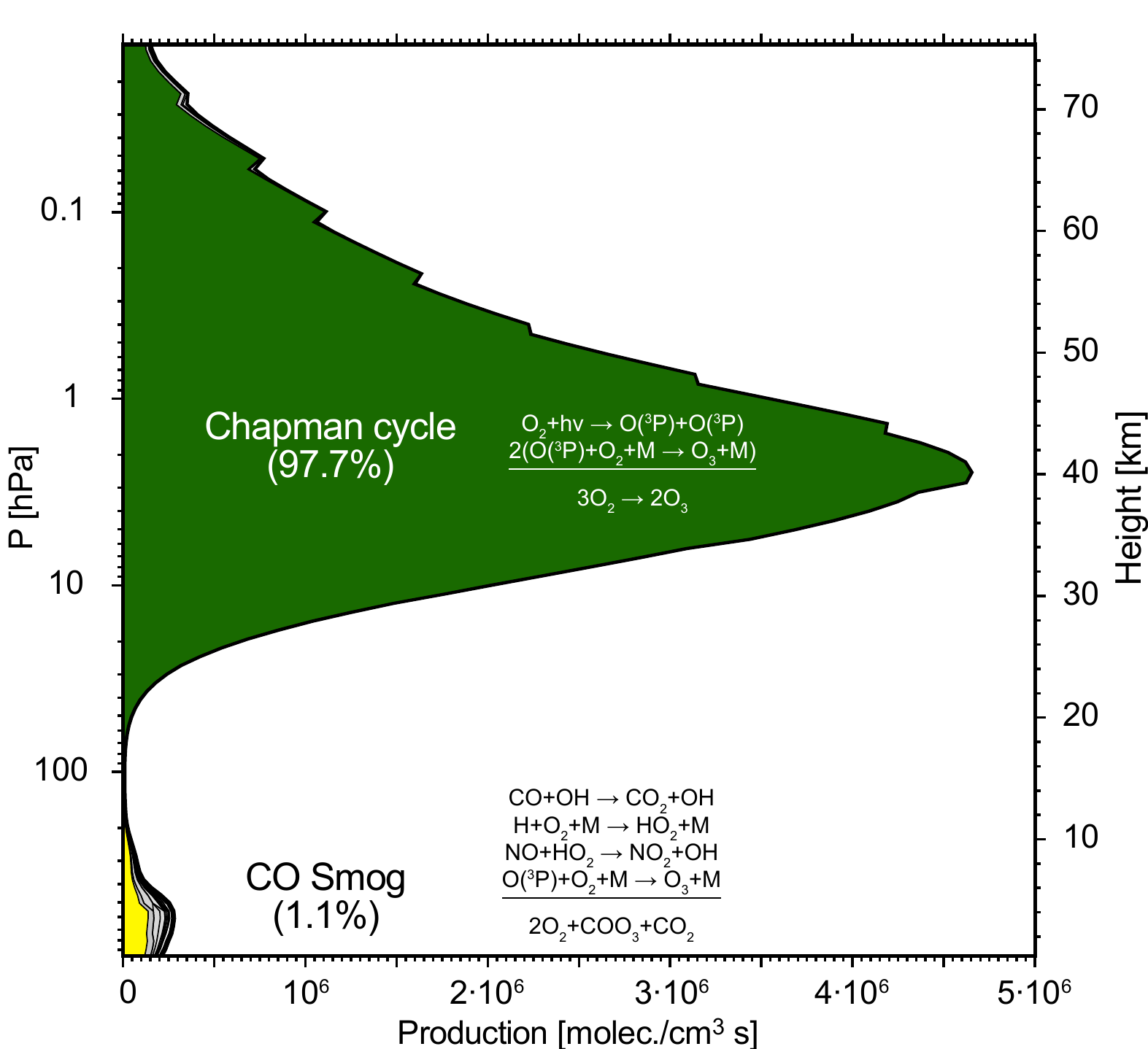}
 \includegraphics[width=0.49\textwidth]{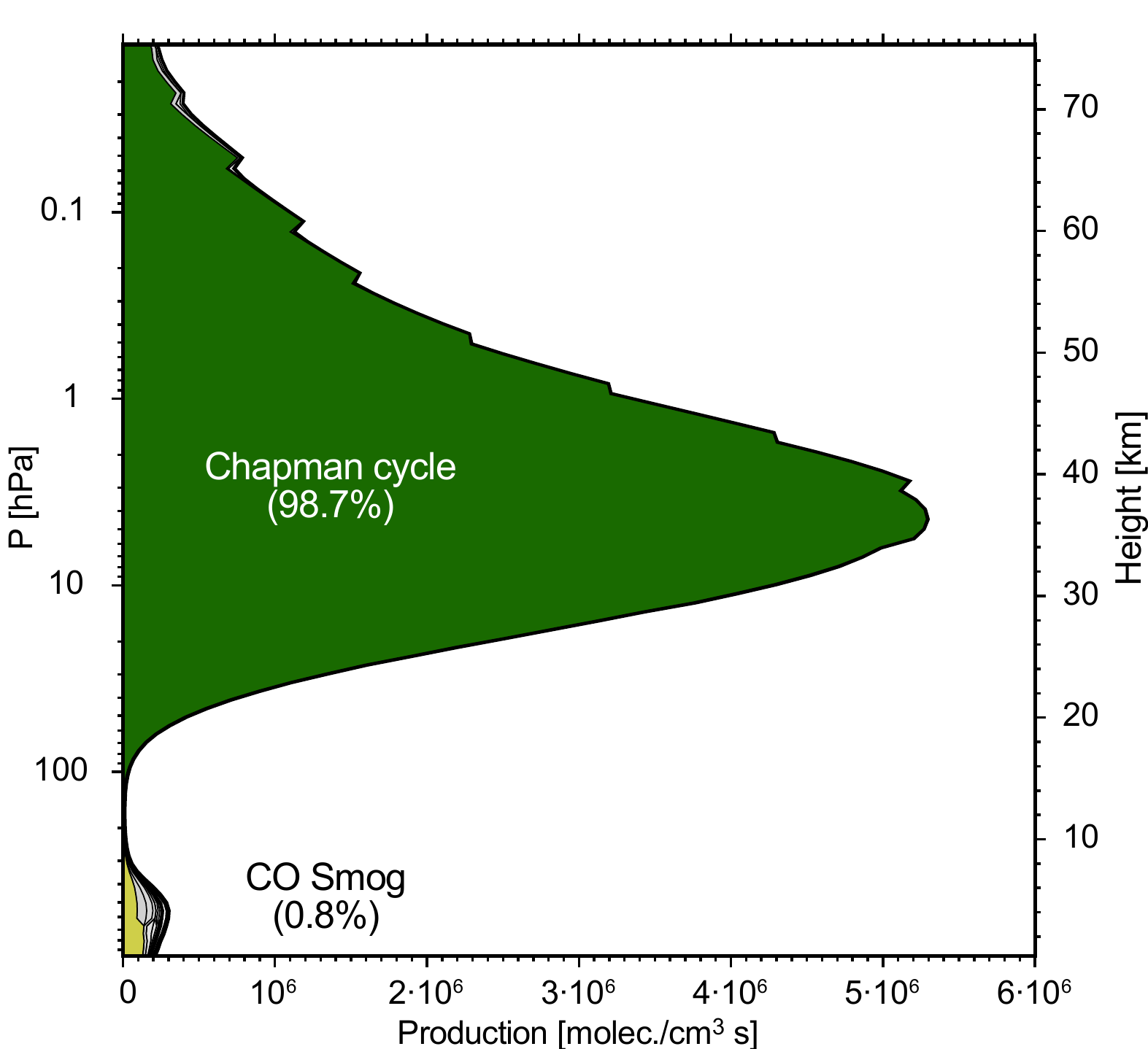}\\
  \caption{Left panel: Ozone production cycles determined by PAP for scenario 1 (quiescent Sun). Right panel: Ozone production cycles determined by PAP for scenario 2 (GLE05).}
 \label{fig:11_1}
 \end{figure*}
%
\begin{figure}
\begin{center}
 \includegraphics[width=0.7\columnwidth]{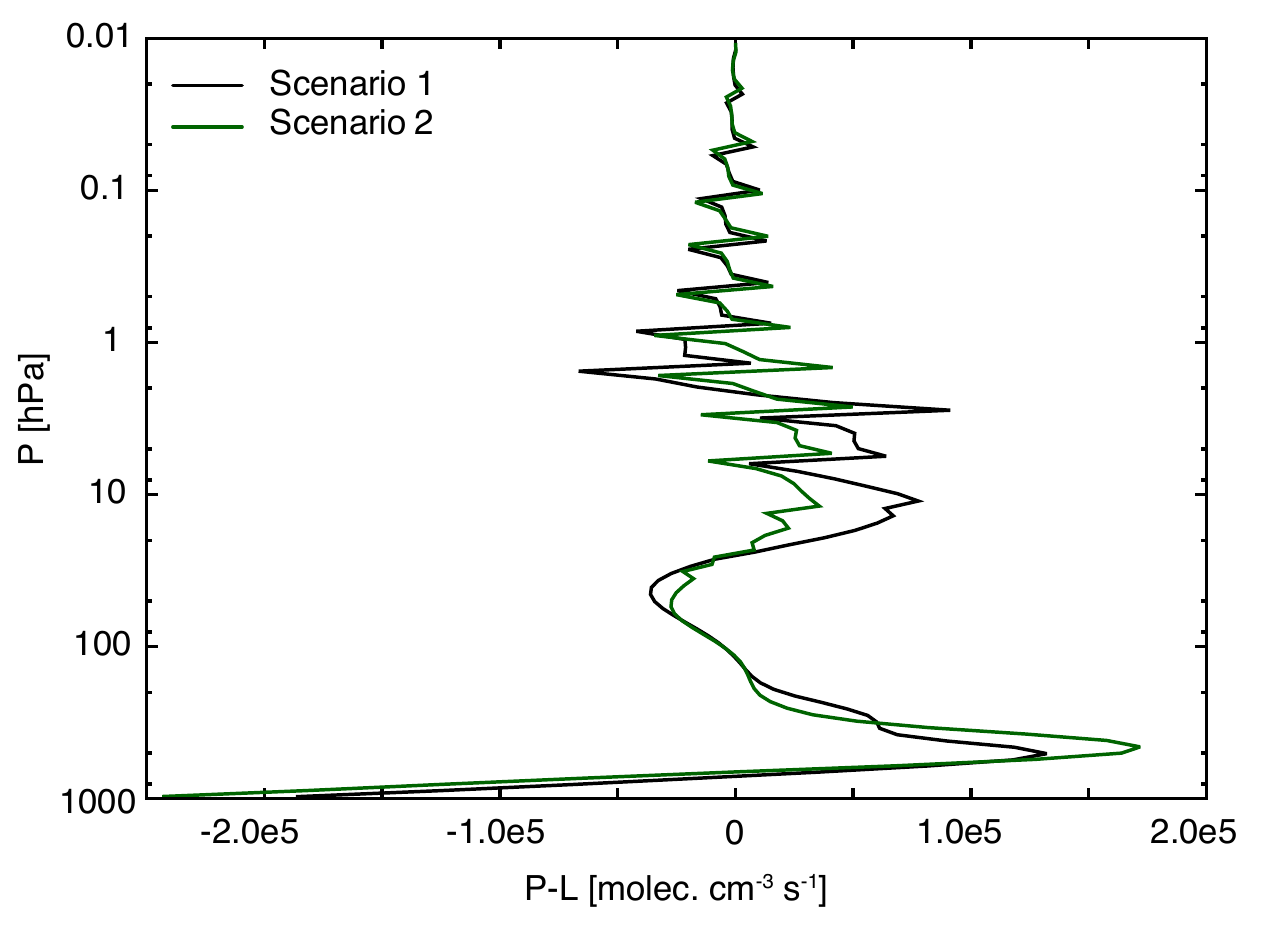}
  \caption{ Production (P) minus Loss (L) of gas-phase ozone (O$_3$) for scenario 1 (quiescent Sun, black line) and scenario 2 (GLE05, green line) versus pressure (hPa).}
 \label{fig:11_2}
 \end{center}
 \end{figure}
%
\begin{center}
\begin{figure*}[!t]
  \includegraphics[width=\textwidth]{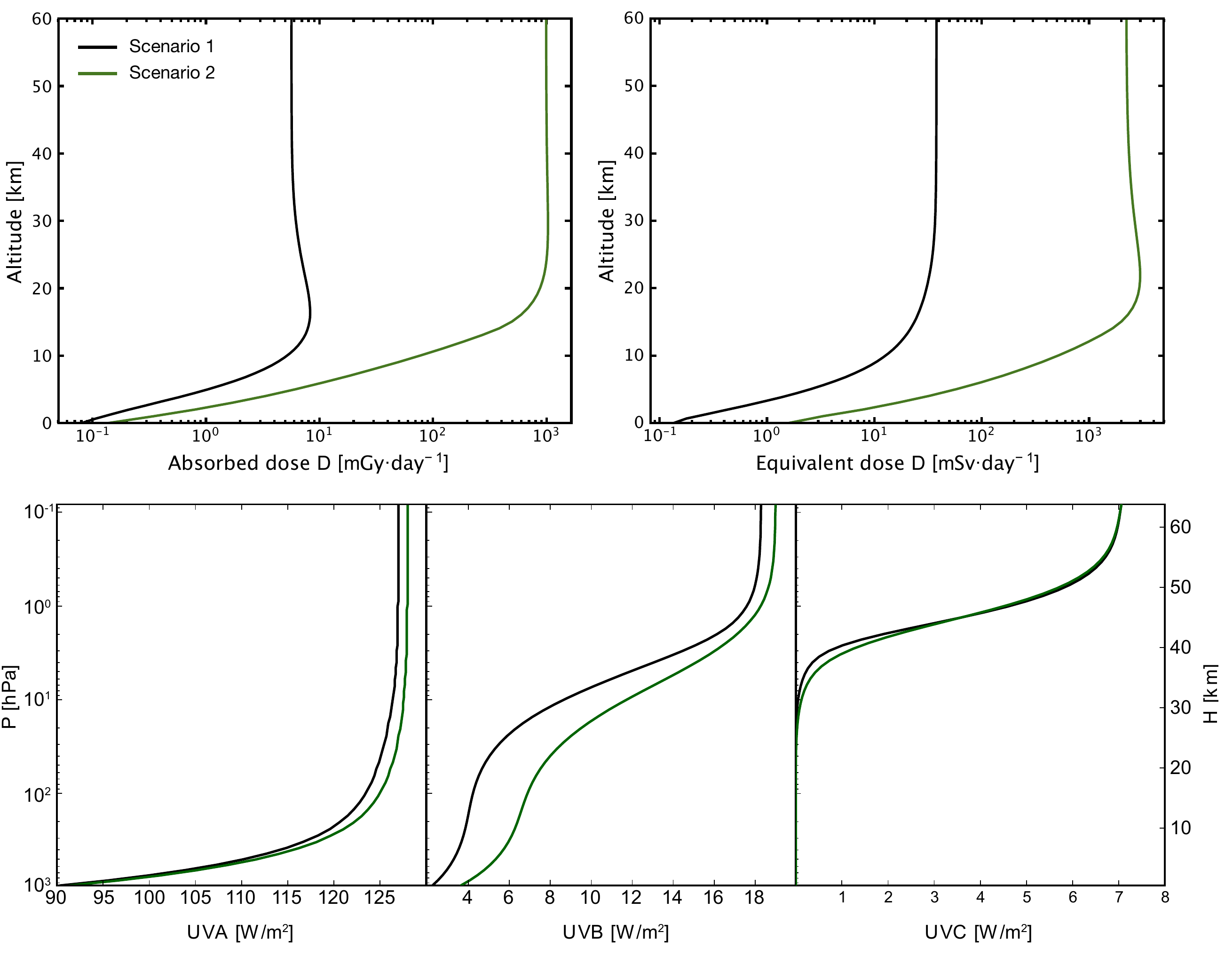}
  \caption{Upper panels: Computed absorbed (left panel) and equivalent (right panel) dose rates during scenario 1 (quiescent Sun, black lines) and scenario 2 (GLE05, green lines). Lower panels: Atmospheric profiles of the UVA (left), UVB (middle), and UVC (right) environments according to both scenarios. }
 \label{fig:9}
 \end{figure*}
\end{center}
\subsection{How do CRs and the resulting photo-chemistry affect the abundance of potential biosignatures?}

The panels of Fig.~\ref{fig:11} and Fig.~\ref{fig:11_1} show the contribution of ozone loss cycles and ozone production cycles for scenario 1 (left panels) as well as for scenario 2 (right panels) as a function of atmospheric pressure and altitude, respectively. These results were obtained with the Pathway Analysis Program \citep[PAP, ][]{Lehmann-2004} which analyses the chemical output from the 1D-TUB model.

The results in Fig.~\ref{fig:11} illustrate the well-studied ozone-UV feedback, a “self-healing” mechanism \citep[negative feedback, see, e.g.,][]{WMO-1994} which helps to maintain the ozone layer and operates as follows: processes which reduce ozone, for example, due to NO$_x$ and HO$_x$ production from cosmic rays lead to stronger penetration of UV into the layers immediately below. Here, they result in stimulated oxygen photolysis, hence enhanced ozone formation. 

As can be seen, the peak loss rates for ozone pathways increase by ~15\% i.e. from 4.6$\cdot$10$^6$ molecules cm$^{-3}$ s$^{-1}$ in scenario 1 up to 5.3$\cdot$10$^6$ molecules cm$^{-3}$ s$^{-1}$ in scenario 2. Additionally, the altitude corresponding to the peak loss rates moves downwards from ~43.5km in scenario 1 to ~40.5km in scenario 2. For the quiescent scenario NO$_x$ cycles are important in the low to mid stratosphere, whereas HO$_x$ and O$_x$ cycles dominate the mid to upper stratosphere. In scenario 2 three NO$_x$-cycles are stimulated by the interacting cosmic rays and they dominate NO$_x$-loss pathways for ozone in the low to mid stratosphere. Note that some lines show a sawtooth-type pattern especially in the upper atmosphere which is related to an interpolation between the different vertical grids of the climate and chemistry modules.

The apparent increase of NO$_x$ in the right compared with the left panel leads to a reduction of ozone (as can also be seen in Fig.~\ref{fig:8}). Therefore, deeper penetration of UV occurs, which shifts the maximum of the ozone production to lower altitudes. The ozone production by the 'CO smog cycle' in the troposphere also increases by a factor of two during the flaring scenario. 

Note that we required PAP to connect partial pathways producing O($^3$P) with partial pathways consuming O($^3$P) at all altitudes i.e. including above 60~km although the lifetime of O($^3$P) is longer than that of ozone there. Without this specification, the PAP does not build a Chapman-pathway above ~60km and instead outputs the single reaction: O($^3$P) + O$_2$ + M $\rightarrow$ O$_3$ + M as the main  O$_3$ source \citep[see, e.g., Figure 6a in][]{Grenfell-etal-2013}.

Figure~\ref{fig:11_2} shows the net production (production minus loss, P-L) of ozone due to the changes in atmospheric chemistry. The results suggest that scenario 2 (GLE case, green line) is shifted to the left of the modern quiescent Earth case (black line) in the heart of the ozone layer at around 10 hPa ($\sim$ 30 km). This is consistent with the ozone loss due to NO$_x$, O$_x$ and HO$_x$ cycles in scenario 2 displayed in Fig.~\ref{fig:11}. In the upper troposphere between 200 and 500 hPa, the results of scenario 2 (green line) are shifted to the right compared with the results of scenario 1, which is consistent with an ozone production due to the smog cycle (favoured by increased NO$_x$).

The rates of the NO$_x$ (= NO + NO$_2$) cycles presented in Fig.~\ref{fig:11} show the combined effect of increased NO$_x$ and decreased ozone concentrations (scenario 2 versus scenario 1). In order to demonstrate the effect of changes of NO$_x$ on ozone more clearly, we calculate pseudo first-order loss rate coefficients of ozone. For this, let us assume that some process, for example a catalytic cycle, removes ozone at a rate $R$:
\begin{equation}
\frac{\delta [\mathrm{O}_3]}{\delta t} = - R.
\end{equation}
A simple reformulation of this equation yields: 
\begin{equation}
\frac{\delta[\mathrm{O}_3]}{\delta t} = -K \cdot [\mathrm{O}_3],
\end{equation}
where the pseudo first-order loss rate coefficient
\begin{equation}
K = \frac{R}{[\mathrm{O}_3]}
\end{equation}
expresses how fast ozone is removed by the process under consideration. Table \ref{tada} shows the pseudo first-order loss rate coefficient of ozone with respect to the first NO$_x$ cycle (NO$_x$1) at 34.6~km where it has its maximum contribution (Figure 8). The scaled rates in Table \ref{tada} (which are taken from the PAP analysis to be the mass flux through the NO$_x$ 1 cycle) suggest that this cycle destroys ozone faster (in this case by a factor 3.2) in scenario 2 than in scenario 1. Similar results have been found for the other NO$_x$ cycles (not shown here).

\begin{table}
\centering
\begin{tabular}{c  c c c}       
\hline\hline                      
Scenario & $\rho_{\mathrm{total}}$ [molec/cm$^3$] & O$_3$ (ppm) & Scaled rate\\
\hline
1 & 1.73$\times 10^{17}$ & 9.44 & 9.05$\times 10^{-7}$\\
2 & 1.71$\times 10^{17}$ & 5.01 & 2.86$\times 10^{-6}$
\end{tabular}
\caption{Pseudo first-order loss rate for both scenarios at 34.6 km (see text). Here, $\rho_{\mathrm{total}}$ represents the total air density and the scaled rates are given by $R$/(O$_3\cdot \rho_{\mathrm{total}}$).}  
\label{tada}
\end{table}
%
\begin{center}
\begin{figure*}
 \includegraphics[width=\textwidth]{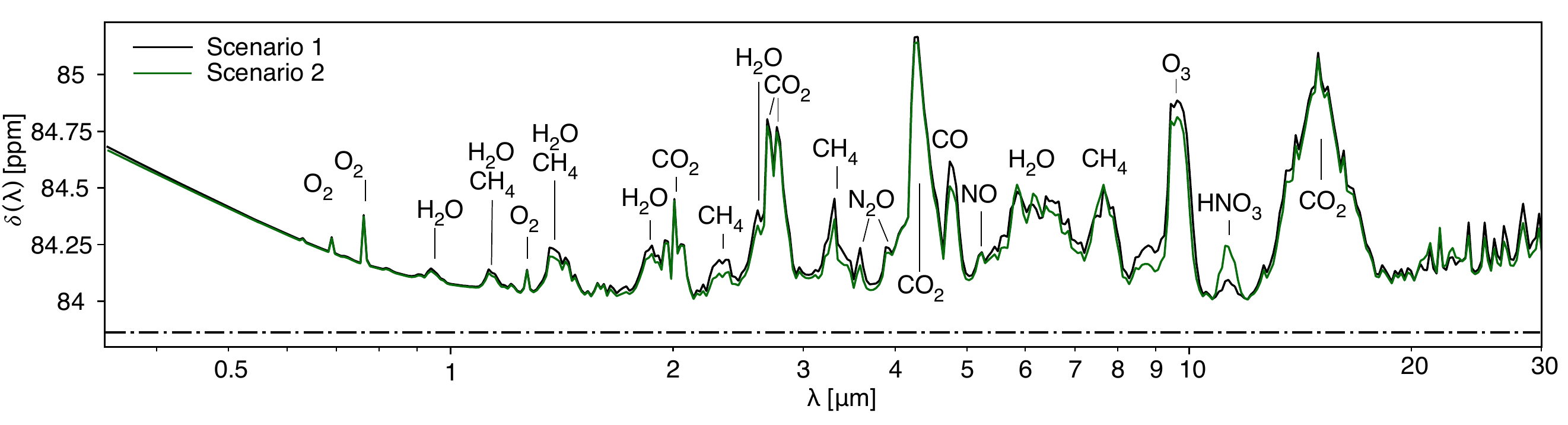}
  \caption{Modeled transit depths in parts per million over wavelength $\lambda$ (R=100) for Earth around our G-type star. Shown are the results of scenario 1 (quiescent Sun, in black) and scenario 2 (GLE05, in green). The dashed-dotted black line at the bottom represents the transit depth for Earth without atmosphere (83.7 ppm).}
 \label{fig:spectra}
 \end{figure*}
\end{center}
%
\begin{center}
\begin{figure}
 \includegraphics[width=\columnwidth]{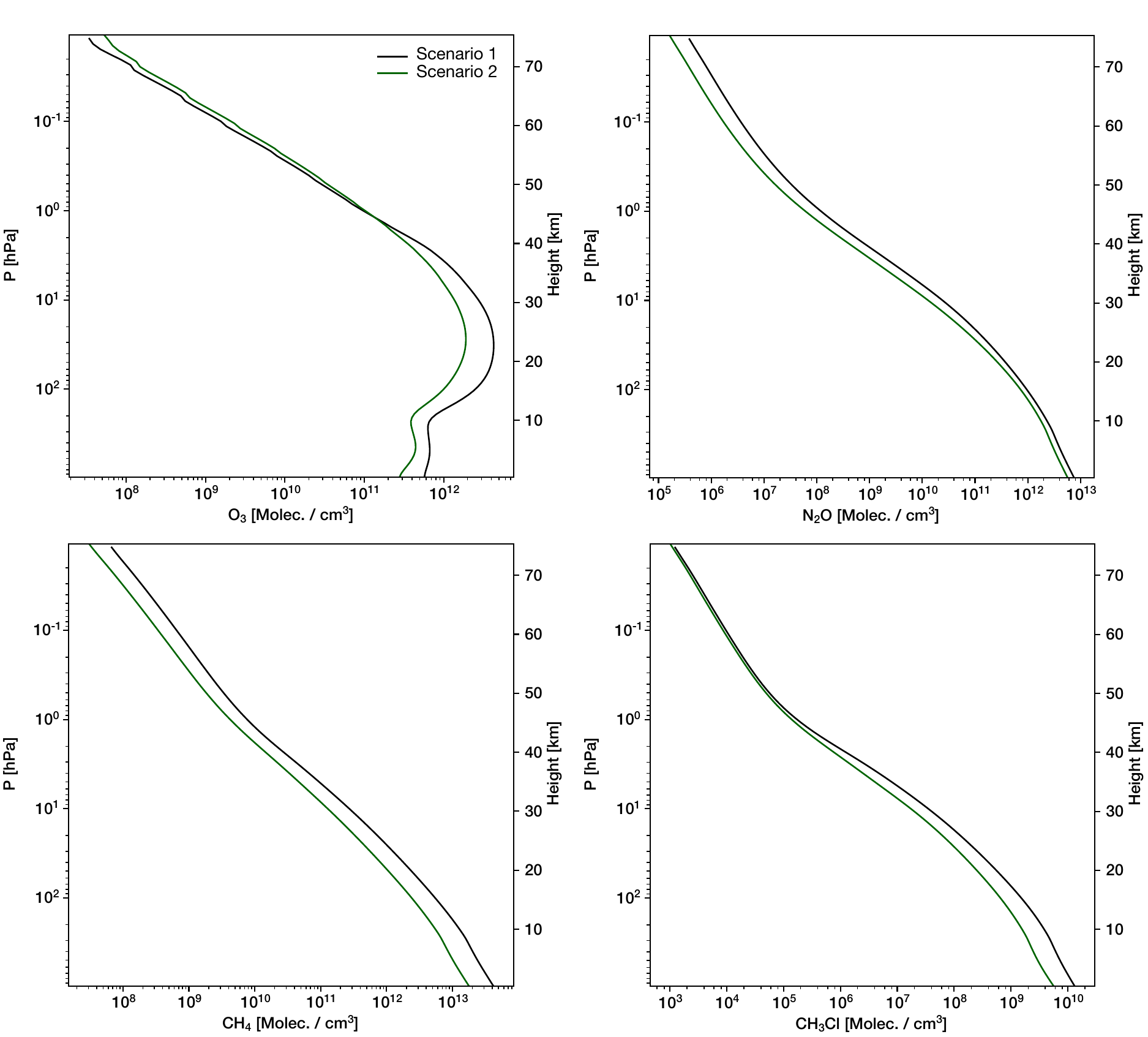}
  \caption{Atmospheric profiles of the atmospheric biosignatures commonly discussed in exoplanetary literature: O$_3$ (upper left panel), N$_2$O (upper right panel), CH$_4$ (lower left panel) and CH$_3$Cl (lower right panel). Shown are the results of scenario 1 (quiescent Sun, black lines) and scenario 2 (GLE05, green lines).}
 \label{fig:8}
 \end{figure}
\end{center}
\subsection{What is the radiation dose at the surface of (exo)planet Earth and how strong is the surface UV radiation exposure?}
Strong solar particle events, in particular those that alter the radiation field near the planetary surface (GLEs), lead to an additional ionization as well as production of secondary particles throughout the entire atmosphere (see also Fig.~\ref{fig:6}). Most recently, a new feature of \ac{AtRIS} was developed, which, for the first time, directly enables parametric studies of the interrelation of incident radiation and the resulting absorbed dose throughout the atmosphere \citep[see][]{banjac2018b}. Utilizing their approach, we computed the altitude-dependent absorbed and equivalent dose rates for both scenarios. The results of this study are shown in the upper panels of Fig.~\ref{fig:9}. Here the absorbed dose rates are displayed on the left, while the equivalent dose rates are given in the right panel. As can be seen, a strong GLE event (green lines) such as the one on February 23, 1956 has a strong impact on both quantities throughout all atmospheric layers. Thereby, for example, the surface absorbed dose rate increases by a factor of two, while at altitudes above 20~km a relative increase of almost two orders of magnitude becomes evident.

Moreover, the lower panels of Fig.~\ref{fig:9} show the effect of the CR-induced chemical changes on the atmospheric UV environment, which are a direct output of the 1D-TUB model. As can be seen, neither UVA nor UVC show any significant differences between the quiescent case with GCR background (black) and the flaring case (green). The UVB environment, on the contrary, shows a clear increase in the lower atmosphere. In our Earth-like atmosphere, the main absorber in the UV-B range is O$_3$. The enhanced UVB environment shown here corresponds well with the reduction of O$_3$ by up to 50\% for the flaring case, as shown in Figs.~\ref{fig:spectra} and \ref{fig:8}. 

As a direct consequence, during a solar event as strong as the February 1956 event a harmful UVB radiation of about 3.5 W/m$^2$ would result at the planetary surface \citep[see also][and references therein for discussion on influence of surface UV-B upon biogeochemical feedbacks during Earth’s evolution]{Gebauer-etal-2017}.
\subsection{How do CRs affect the spectral features of biosignatures?}
Figure~\ref{fig:spectra} shows the modeled transmission spectra derived with the GARLIC model for the two scenarios based on the final-iteration input of (T, p, composition) from the 1D-TUB model. The calculated wavelength range extends from 0.3 to 30$\mu$m, corresponding to one of the modes of the NIR-Spec infrared spectrometer aboard the upcoming JWST mission. Characteristic spectral features of the modern Earth are seen, for example, the Rayleigh extinction 'slope' in the UV-visible, the rotational-vibrational water and methane bands in the near IR and the characteristic ozone and carbon dioxide fundamentals at 9.6$\mu$m and 15$\mu$m, respectively. Weaker spectral features of proposed atmospheric biosignatures such as O$_2$ (in the visible) and N$_2$O (at around 7.8$\mu$m) are also apparent. A comparison between the two scenarios shows a moderate suppression of some spectral features for the GLE event (green lines), e.g., for O$_3$ (9.6$\mu$m) and CH$_4$ (near IR) due to, e.g., photochemical loss induced by cosmic rays. In addition, \citet{Tabataba-Vakili-2016} proposed an HNO$_3$ feature around 12 $\mu$m as a cosmic-ray signature for Earth-like planets around active M-stars. The fact that we see the same feature during a strong GLE in the Earth's atmosphere (see Fig.~\ref{fig:spectra}) strengthens their argument. However, \citet{Scheucher18} recently showed that these results strongly depend on the strength of the event as well as the atmospheric composition of the exoplanetary atmosphere.

Although the GLE event of February 1956 was rather modest compared to what has been observed for flaring M-dwarfs like AD Leonis or Proxima Centari \citep[see, e.g.,][]{Herbst-etal-2019a}, its impact on the atmospheric biosignatures is clearly visible. For a better overview of the differences in the transit spectra Fig.~\ref{fig:8} displays the abundance changes for the molecules O$_3$ (upper left panel), N$_2$O (upper right panel), CH$_4$ (lower left panel) and CH$_3$Cl (lower right panel): These molecules, in particular, are known on Earth to be heavily influenced by life. In the case of O$_3$, our results show decreases of up to 50\% in the lower-mid stratosphere as well as the lower troposphere for the GLE event of 1956 (green line). Even though we assume these high particle fluxes to be constant in our steady-state assumption, this result compares well with the measured O$_3$ depletion after such big events in higher latitudes for a few months \citep[see, e.g.,][]{McConnell-etal-2017}. The effect of GLE05 on other molecules that are not shown here is, however, overall, only minor throughout our model atmosphere.
\section{Summary and Conclusions}

Here we presented the validation of our self-consistent model chain to determine the influence of cosmic rays on (exo)planetary atmospheric biosignatures based on the example of Earth as an exoplanet in the habitable zone of a G-type star.

Thereby, we focused on the modeling of the impact of the stellar radiation and particle environment on exoplanetary atmospheric climate and chemistry, as well as on potential atmospheric biosignatures for the modern Earth as an example of an Earth-like exoplanet in the habitable zone of a G-type star. Therefore, the numerical models developed and/or updated at the three institutions were used in order to build a self-consistent simulation chain by taking into account the GCR and SCR flux, the particle transport and interaction within the planetary magnetic field (utilizing PLANETOCOSMICS), the atmosphere by producing secondary particle cascades, and thus atmospheric ionization and radiation (utilizing AtRIS), as well as the cosmic-ray induced influence on the atmospheric chemistry (1D-TUB) and the ion-chemistry (ExoTIC). 

Energetic particles are known to have a strong influence on the atmospheric ion-chemistry, and we showed that it is mandatory to iterate between the different models in order to assure convergence between the different codes. While for GCRs only one iteration between the different models was necessary, in case of the strongest GLE event ever measured (GLE05) at least three iteration rounds were needed until the modeled ionization rates converged (see Fig.~\ref{fig:5}). We showed that the GCR-induced atmospheric ionization is roughly two orders of magnitude lower at the Pfotzer maximum (16-20 km, see Fig.~\ref{fig:6}) than the SCR-induced values. As a consequence, also the reduction in the production rates of, for example, NO, H or OH due to the induced ion-chemistry processes is in the same order (see Fig.~\ref{fig:chemistry}).

Based on these findings, we computed the influence of the induced photochemical processes on potential biosignatures. Utilizing an algorithm for the determination of chemical pathways (PAP) the contribution of ozone loss cycles (see Fig.~\ref{fig:11}) as well as sources (see Fig.~\ref{fig:11_1}) were studied. The results are in agreement with the well-studied ozone-UV feedback \citep[negative feedback, see, e.g.,][]{WMO-1994} which helps maintain the ozone layer.

Moreover, the altitude-dependent absorbed and equivalent dose rates, the atmospheric UVA, UVB, and UVC profiles for both scenarios were computed. We showed that a strong GLE such as the one on February 23, 1956 not only has a substantial impact on, for example, the surface absorbed dose rate, which also increased by up to two orders of magnitude, but also the UVB flux, which would result in a harmful flux of about 3.5 W/m$^2$ at the planetary surface. 

Although the studied GLE event was rather modest compared to what has been observed for other Sun-like stars \citep[see][]{Notsu-etal-2019} and flaring M-dwarfs like Ad Leonis or Proxima Centauri \citep[see, e.g.,][]{Herbst-etal-2019a}, its impact on the atmospheric biosignatures is clearly visible (see Figs.~\ref{fig:spectra} and \ref{fig:8}). Thereby, for example, O$_3$ is decreased by up to 50\% in the lower-mid stratosphere as well as the lower troposphere due to the influence of this strong SEP event.
Taking into account that old Sun-like stars have been found to produce superflares with energies above 5$\cdot 10^{34}$erg once every 2000 to 3000 years \citep[see][]{Notsu-etal-2019} it is fair to assume that stronger GLE events may occur on the Sun. Imprints of such events have already been found in the cosmogenic radionuclide records of $^{10}$Be, $^{14}$C, and $^{36}$Cl around AD774/5, AD992/3, and 660 BC \citep[see, e.g.,][ respectively]{Miyake-etal-2012, Mekhaldi-etal-2015, OHare-etal-2019}. Since such energetic flares could result in proton events about 200 times stronger than GLE05 \citep{Herbst-etal-2019a}, it is highly likely, that the atmospheric O$_3$ depletion would be much stronger. Even stronger flares with energies up to 10$^{36}$ erg (10 W/m$^2$) have been detected on K- and M-stars \citep[see, e.g.,][]{Candelaresi-etal-2014, Howard-etal-2018}. Such flares could result in particle events more than five orders of magnitude stronger than GLE05 \citep{Herbst-etal-2019a} which could drastically impact on planetary habitability. 

These results indicate that ozone is likely depleted in N$_2$-O$_2$-dominated atmospheres of planets in the habitable zones of frequently flaring stars. Furthermore, due to the slow recovery of ozone after flare events, ozone might not be a good biomarker for planets orbiting stars with high flaring rates.

Initial model efforts to study the impact of such superflares on planetary habitability have been recently published by \citet{Yamashiki-etal-2019} and \citet{Airapetian-etal-2019}. In addition, the newly-selected ISSI Team "The Role Of Solar And Stellar Energetic Particles On (Exo)Planetary Habitability (ETERNAL)", lead by KH and JLG, will bring together scientific expertise from the fields of solar-, planetary and exoplanetary science in order to determine the impact of the stellar particle and radiation field around main sequence stars on the atmospheric chemistry, climate and habitability of exoplanets with N$_2$-O$_2$-, CO$_2$-, H$_2$-, and H$_2$O-dominated atmospheres in the HZ of their host stars.
\begin{acknowledgements}
We thank the German Research Foundation (DFG) for financial support via the project \textit{The Influence of Cosmic Rays on Exoplanetary Atmospheric Biosignatures} (Project number 282759267). KH, JLG, and BH acknowledge the International Space Science Institute and the supported International Team 464: The Role Of Solar And Stellar Energetic Particles On (Exo)Planetary Habitability (ETERNAL). KH and BH further acknowledge the supported International Team 441: High EneRgy sOlar partICle Events Analysis (HEROIC). FS is supported by the DFG project SCHR1125/3-1.
\end{acknowledgements}
\bibliographystyle{aa}
\bibliography{references}
\end{document}